# Everything Counts: The Managed Omnirelevance of Speech in Human–Voice Agent Interaction

The omnirelevance of speech in 'human – voice agent' interaction

Damien Rudaz

University of Copenhagen, daru@hum.ku.dk

Mathias Broth

Linköping University, mathias.broth@liu.se

Jakub Mlynář

HES-SO Valais-Wallis, jakub.mlynar@hevs.ch

To this day, turn-taking models determining voice agents' conduct have been examined primarily from a technical point of view, while the ways in which they emerge as interactional constraints or resources for human conversationalists in situ remain underexplored. Drawing on a detailed analysis of corpora of naturalistic data, we document how humans' conduct was produced in reference to the ever-present risk that, each time they spoke, their talk might trigger a new uncalled-for contribution from the artificial agent. We examine this phenomenon in interactions involving rule-based robots from a 'pre-LLM era' as well as the most recent voice agents. This 'omnirelevance of human speech' (i.e., the possibility that a conversational agent may erroneously respond to any speech it detects) emerged as a constitutive feature of these human-agent encounters. We describe some of the practices through which humans managed these artificial agents' turn-taking conduct. Given recent improvements in voice capture technology, we ask whether this 'omnirelevance of human speech' weighs even more heavily on human practices today than in the past.

# 1 INTRODUCTION

## 1.1 Documenting the interactional relevance of turn-taking models

This study investigates how humans organize their talk in the vicinity of recent conversational voice agents. In doing so, it describes the ways in which conversational agents' turn-taking models *emerge* as relevant in naturalistic interaction. Relying on the detailed analysis of situations involving humans and artificial agents, we find that, whether in exchanges with rule-based robots or recent LLM-based voice agents[1], human activity was observably shaped with reference to a constant, looming constraint. This Sword of Damocles, hanging over every turn-at-talk, was the constant possibility that co-present artificial agents might respond to any hearable speech as projecting a next action on their part [69, 126]. We report how humans practically handled the ever-present risk that, each time they spoke, their talk might trigger a new contribution from the artificial agent. In this way, we explore how prevailing properties of some turn-taking models (those currently used by the overwhelming majority of voice agents or robots) are oriented to and managed as resources, affordances, or constraints by human participants in situated interaction.

In parallel, this analytic angle offers the opportunity to reconsider whether, with regard to turn-taking, humans' "work to make technology work" [45] has shifted over the past years. We examine whether, behind the immense improvements of contemporary conversational agents, the same typical interactional practices are still produced by human participants – practices which were already characteristic of what 'conversing with an AI' was several years ago. Do human conversationalists rely on distinctly new methods when managing voice-based artificial agents' increasingly refined ability to pick up speech and produce responses? Or, from simplistic 'social' robots to today's most sophisticated conversational systems, *can we rather observe a praxeological continuity* (i.e., the same practices being deployed by users) *despite a technological rupture?*

## 1.2 A pivot point between an interactional perspective and an engineering perspective

The following study rests on a tension between a technical perspective on 'what is going on behind the curtain' as voice agents respond to humans, and an interactional, emic, perspective on what properties of those voice agents emerge[2] as relevant properties of the setting for human conversationalists in situ. Indeed, publicly available conversational AIs have benefited from substantial advances when it comes to composing relevant responses in written form – they "know what to say" [164] – whereas they have made significantly less progress in taking turns during spoken conversation – they do not know "when to talk" [164]. Otherwise put, *turn positioning* (when a spoken turn occurs in interaction) has lagged behind *turn composition* (how a turn is designed – lexically, syntactically, prosodically). The previous formulation subsumes an interactional perspective on what can also be framed from a technical point of view: at the time of writing, the most widely used recent conversational agents rely on simple silence-based turn-taking models [34, 149, 150, 154] rather than complex continuous turn-taking models [62, 63, 149] or full-duplex speech-to-speech systems [128]. This technological backdrop determines the patterns of conduct of voice agents – for example, erroneously responding to continuers like "uh huh" [139] – that are locally encountered as constraints or resources by users as they interact with those agents. We detail how co-present humans enact various practices through which they progress the talk with or around those artificial agents, whether those practices are routinized or, conversely, improvised 'on the spot' after extended periods of friction with these agents. We do so using two types of video data of naturalistic interactions:

1. Excerpts collected from naturalistic interactions between visitors and the 'social' robot Pepper in a museum, in 2022. These data were collected before LLM-based conversational systems were made available to the general public. They represent a "pre-LLM era" [71], when most conversational agents (including the Pepper robot in these videos) relied on more or less complex conversational trees [18, 82] as part of conventional rule-based dialog systems [170].
2. Excerpts collected from naturalistic interactions with OpenAI's recent multimodal "advanced voice mode", in 2025. These interactions feature human participants talking to this voice agent on their smartphone, as part of other encompassing activities (here, playing cards and reading the newspaper at home).

---

1 We define voice agents (also referred to as voice-based agents [40, 132]) as a particular subset of conversational agents – see [19, 70] for a similar use of the term. The term conversational agent generally refers to systems capable of engaging in natural-language dialogue through text, speech, or other modalities, while a voice agent denotes a conversational agent whose interaction with users is mediated exclusively through spoken language [166].

2 The term emergence refers here to a situated accomplishment of participants [96, 101, 124]. It is through participants' conduct that some parameters of a setting (among all the features of this setting that can be segmented) become relevant within the ongoing activity.

In doing so, our study contributes to the field of Human–Computer Interaction (hereafter HCI) by exploring the under-documented interactional counterpart to the well-documented technical features of current turn-taking models. This analytic perspective is 'technosolutionism-agnostic' (see Section 7.2): it studies technology as it is currently used, regardless of 'what may soon be solved' by advances in generative AI more broadly or in systems governing artificial agents' turn-taking behavior in particular.

## 2 RELATED WORK

### 2.1 Turn-taking models and their limits

A wide range of research has approached the issue of creating conversational agents that are able to take turns relevantly in natural spoken conversation [62, 63, 104, 149, 157]. A core difficulty within this endeavour is to be found in the transition from *normative* turn-taking rules identified by Sacks et al. [137] to *computational* processes likely to enable conversational agents to act in accordance with these normative rules [149, 169] in situated interaction. The central tension lies between, on the one hand, describing the "formal structures of practical action" [39] – e.g., the organization of conversation "that stands apart from any particular actor" [17] – and, on the other hand, designing the inner workings of artificial agents (their software and hardware) so as to allow those agents to re-accomplish those formal structures within the contingencies of each new situation. Indeed, mapping the interpersonal conversational "machinery" [136] – both oriented to and co-produced by participants in a setting – does not provide engineers with insight into the internal 'machinery' of each actor (human or artificial) that enables them to take part in this joint accomplishment of a locally organized activity[3].

In this endeavour to make a robot or a voice agent's conduct "consistent with a number of observations and constraints" [79] (those documented many years ago by Sacks et al. [137]), simple silence-based turn-taking models [149][4] are commonly used as a convenient but ultimately limited crutch. Moreover, these silence-based turn-taking models often rely on simple rule-based systems: they regularly (although not necessarily) consist of a few 'if–then' rules. In such cases, a consequential discrepancy emerges between "the sensitivity of human participants to the normative order of conversation" [86] and a fundamental insensitivity to such an order on the part of voice agents relying on these basic turn-taking models. Whereas humans orient to norms, contextual gestalts [86], and frameworks (connected to, e.g., specific institutional

3 Within the field of ethnomethodology and conversation analysis, a description of this fundamental gap (between describing normative rules and designing the inner-workings of an agent whose conduct aligns with those rules) can be found in Rudaz [129]. Of course, this practical issue is merely the surface expression of the common sense problem in AI, documented for decades across various intellectual traditions [21, 25, 91].
4 See section 3.3 for a contrastive definition of the terms turn-taking systems and turn-taking models.

contexts [54]) as resources in taking turns, simple silence-based models do not allow artificial agents to operate in the same way. This is where a breakdown occurs between insights from conversational analytic literature and the opportunities offered by the technical infrastructure of these artificial agents: those agents cannot be designed to 'orient' to the normative organization of conversation as a local resource among any other [16, 36, 161] in accomplishing specific actions or, more generally, "in doing social life" [135].

To mitigate the previous shortcomings, machine-learning-based full-duplex spoken dialogue models [1, 24, 128] or continuous turn-taking models [61, 62, 77, 110, 150, 157] have been developed. Rather than relying on fixed rules – such as waiting for a predetermined silence duration to recognize the end of a turn – these systems dynamically predict turn transitions at any point in the unfolding speech stream. Many continuous models now integrate prosodic [31], syntactic, semantic, pragmatic [150], and even visual dimensions [74, 110] to turn-taking. These data-driven approaches have been shown to enable smoother exchanges [150], more attuned to the subtle cues and timing of human turn-taking. Yet, such systems remain almost entirely absent from today's commercial voice-user interfaces and social robots [34, 154], which continue to rely on coarse silence-threshold or basic voice activity detection heuristics to detect turn completions.

### 2.2 Studies of robots and voice agents *in use*

A significant body of work has approached turn-taking in its concrete, situated, accomplishment in settings involving artificial agents [33, 57, 86, 113, 117, 120, 158]. These studies focus on robots and voice agents *in use* by attending to the sequential and multimodal details of activities involving such agents (for an overview, see [93]). They have repeatedly documented how, in everyday settings, voice interfaces (such as Amazon Echo [33, 120]) or 'social' robots [73, 86] fail to take or to allocate turns – and how human participants repair such breakdowns. These works reveal that some of the most intricate local contingencies artificial agents must contend with emerge from the number of participants present in a given interaction. Indeed, robots and voice agents are usually designed for dyadic interactions with a single user [57]. Yet, in real-world settings, interactions involving these agents are frequently multi-party [121]; in these conditions, the accountable absence of fine-tuned adjustments from the machine is made especially apparent. For instance, Majlesi et al. [86] report how human participants managed problems of turn-taking in multiparty interactions with robots (e.g., through restarts in case of overlaps). They describe how the robots' limited orientation to the sequential and multimodal context was, in the end, accommodated by humans. Similarly, Jarske et al. [65] analyse multiparty interactions with the social robot Pepper, highlighting how participants collaboratively adapted their speech and its timing – termed "robot speak" – to match the robot's rudimentary capabilities to listen and respond.

However, although exacerbated (and rendered observable) by the multi-party nature of some interactions, the "work to make technology work" [45] is not restricted to such configurations. In dyadic interaction, Pelikan and Broth [113] describe how participants adjusted their speech timing and turn design to match the robot's predefined slots for responses – adding to the body of documented cases in which humans perform interactional work to accommodate robots' constraints related to turn-taking. They describe how humans progressively recognized 'when to talk', that is, to produce speech only during the narrow time windows when the NAO robot would be able to hear them. Similarly, Oloff [109] reveals how older adults adapted their turn-taking practices to match voice assistants' 'listening' slots. In doing so, users learning to interact with these systems oriented to the temporal positioning of their talk as a practical problem. These participants had to develop new, sui generis, practices to progress the interaction with voice agents – rather than rely on practices built from and for human conversation. In interacting with a voice agent, what counted as 'adequate' or 'orderly' talk had changed: human users had to structure and design their talk differently.

### 2.3 Computational and experimental studies on the effect of turn-taking conduct

While the aforementioned studies highlight the (endogenous) interactional relevance of turn-taking constraints, their findings are not causally independent from the (exogenous) properties of artificial agents' turn-taking models. However, because these works focus on the emergent categories [94, 99] relevant to participants themselves, they do not aim to document these turn-taking model designs directly, nor to test the interactional or pragmatic consequences of different versions of those models. This task is taken up by other research drawing on different methodological perspectives. These experimental or semi-experimental studies explicitly focus on the effect of different turn-taking models on users' perception of an artificial agent or on their conversational practices [52, 75, 85, 112, 150]. In these works, turn-taking models are explicitly configured as parameters whose impact must be evaluated, regardless of whether the (etic) category of 'turn-taking model' also emerges as relevant for participants in situ.

The vast majority of these studies has examined alternative strategies for artificial agents regarding both turn-taking and turn-design (i.e., how agents should both *position* and *compose* some of their turns [146]). For instance, Żarkowski [171] experimentally compared a basic speech-only turn-taking model with a multimodal system that integrated gaze information – and that "allow[ed] the users to interact with each other after each utterance" rather than speaking after each detected silence. The multimodal approach improved user perceptions – making the robot appear more cooperative, communicative, and intuitive – and enhanced what the authors defined as the fluency of the interaction. Similarly, Ohshima et al. [108] investigated how a robot's behaviour in response to silences (initiating a new turn that specifically addressed an individual, the group, or all participants) shaped participants' conversational practices in group discussions. Additionally, Cumbal et al. [23] compared three interruption-handling strategies for a social agent: ignoring, immediately yielding, and acknowledging interruptions produced by humans. These strategies shaped both users' perceptions of the agent's personality and their own conversational behaviours, such as how long they spoke or how easily they interrupted this agent.

Yet, beyond experimental comparisons of turn-taking strategies, only a limited number of studies has systematically examined how artificial agents' turn-taking practices emerge as troublesome in concrete encounters. Notably, Irfan et al. [64] investigated how older adults perceived and interacted with a companion robot using an LLM-based dialogue system. They identified several "dialogue disruptions" connected to the robot's turn-taking behaviour, including delayed responses and interrupting the user. Similarly, Addlesee & Eshghi's [2] work on making voice assistants dementia-friendly reports that short mid-sentence pauses produced by human users often triggered new turns from the voice assistant. These interruptions led to frustration for the users, as the voice assistant initiated its new turn while a human's turn was still recognizably ongoing.

Nevertheless, even the most detailed of these works do not specify how participants' conduct observably indexes the local constraints introduced by specific turn-taking models or, on the contrary, how these models indirectly provide resources that participants draw upon in situated interaction. In another vocabulary, the previous computational and experimental studies tend to treat interaction as a black box [53]. To fill this gap, in line with a characteristic ethnomethodological stance, the present work seeks to describe 'just how' turn-taking models and their specific features emerge as relevant in naturalistic interactions. We aim to document the interactional counterpart of the technical limitations that shape the turn-taking behaviour of voice agents – even those relying on the most recent language models (see, e.g., [163, 164]).

### 2.4 Silence-based turn-taking models: A brief technical overview

To contextualize the analyses presented in Section 5 below, it is helpful to clarify certain aspects of the turn-taking models employed by the artificial agents on which our corpus is based. A fundamental problem for designers of voice agents is enabling these agents to participate in turn-taking within conversational sequences. To date, a common solution has been for the machine to be sensitive to moments of emerging silence, i.e., to rely on silence-based turn-taking models. The core component of silence-based turn-taking is Voice Activity Detection (VAD). This technology detects when speech is present and when it is absent, with the absence of speech denoted as a silence. When VAD detects a significant period of silence, this information can be used to signal the end of a participant's ongoing turn. For example, by default, the commercial version of the Pepper robot (examined in our first excerpt in Section 5) employs VAD, then "relies purely on silence to detect the end of the user's turn" [158].

Similarly, regarding OpenAI's ChatGPT "advanced voice mode" (featured in the second and third excerpts of Section 5) all evidence points to this mode relying on pauses in speech to determine the boundaries of a turn. Although OpenAI does not disclose the exact turn-taking model used by this "advanced voice mode", substantial confirmation is provided by OpenAI's Realtime API documentation[5]: the Realtime API, which powers ChatGPT's "advanced voice mode", uses a server_vad setting to manage turn-taking. This configuration detects the end of a user's speech based on periods of silence, then prompts the system to generate a response. Accordingly, the turn_detection parameter can be set to server_vad, with adjustable properties such as silence_duration_ms to fine-tune the sensitivity to pauses in speech. This approach aligns with traditional uses of VAD, where the system waits for a period of silence before (systematically) initiating a response [149]. ChatGPT user reports of interruptions due to short pauses also implicitly confirm this silence-based mechanism[6]. Indeed, the reliance on silence-based turn-taking implies that the system determines turn transitions primarily through acoustic cues, rather than semantic understanding. This approach leads the artificial agent to interrupt itself when any hearable speech is detected, as the system will treat any pause after speech as the end of a full-fledged turn (and not, e.g., as a response token [35] such as "uh huh" [139]). This state of affairs is likely to extend to turn-taking systems used by other state-of-the-art conversational agents. For example, Skantze and Irfan [150] note that "[a]lthough the algorithm behind Google's end-of-speech detection is not documented, it is likely based on a silence threshold, since it is fairly constant in length". Such technical features determined some of the patterns of conduct exhibited by these agents that emerged as relevant for participants in the interactions examined in the following sections.

## 3 METHODOLOGICAL APPROACH

### 3.1 An Ethnomethodological and Conversation Analytic approach

To investigate how human conversationalists manage the practicalities [51] of interacting with voice agents, this study adopts an Ethnomethodological Conversation Analysis (EMCA) perspective. Research grounded in EMCA attends to the moment-by-moment production of social order [36, 145]: it examines the sequential [136] and multimodal [43, 49, 98] organization of conduct (including talk, gesture, gaze, and body movement) to understand how participants collaboratively achieve and make sense of social actions in real time. EMCA work typically relies on the detailed analysis of video-recorded naturalistic interactions [48, 50], enabling a fine-grained reconstruction of how participants' conduct (including embodied features such as pointing [95]) is accomplished and calibrated with regard to the specific properties of the setting.

---

5 For a description of VAD within OpenAI's Realtime API, see OpenAI's documentation https://platform.openai.com/docs/guides/realtime-vad (accessed May 22, 2025) or Microsoft's documentation https://learn.microsoft.com/en-us/azure/ai-services/openai/how-to/realtime-audio (accessed May 22, 2025).

6 For an example of such user reports, see e.g., https://community.openai.com/t/feature-request-advanced-voice-mode-keeps-interrupting-me/962909. Accessed May 22, 2025.

This *emic* orientation of EMCA ensures that the analytical focus remains with what participants themselves treat as relevant, rather than what might be assumed to matter by a professional analyst. Rather than imposing external categories or technical assumptions, the analysis focuses exclusively on what participants demonstrably display as perceivable and accountable features of a particular setting through their conduct [90, 105][7]. That is, "[t]he relevance of details is always indexical; it cannot be decided a priori and once and for all" [101].

This methodological stance is particularly suited to studying if and how technological artifacts – such as conversational agents – become consequential in situated interaction [55, 58]. Instead of treating a system's technical features as inherently enabling or constraining, EMCA investigates how these properties are publicly encountered, managed, and made consequential by users in regard to practical, local, problems. Significantly, EMCA practitioners focus on the public accountability of action [145]: the actions of the members of a setting are understood as being produced in a way that makes their meaning and purpose available to co-participants, who in turn display their own understanding in their subsequent actions [36, 76]. This provides a rigorous basis for analysing how, for instance, a user orients to an artificial agent's response as an action of a particular type [69] – a request, a question, a greeting, and so on – regardless of the models or algorithms that may determine the agent's conduct from a computational perspective [30].

In this approach, the lived reality of technological properties is thus studied as a *practical accomplishment* [118]. If exogenous parameters impact the local unfolding of an interaction (including any technical property, algorithm, or language model, that may technically determine the conduct of an artificial agent, but, also, any variable such as the users' age, gender, or 'attitude towards technology'), these parameters will, one way or another, "enter the stream of discourse via particular mechanisms of production" [8]. This claim points to both an observable state of affairs and an epistemological principle: if an 'external' context (e.g., technological) is claimed to be relevant, it must be demonstrated to be so "intra-interactionally" [123, 141][8].

### 3.2 'Conversational agents' as a placeholder category: Analytic rigor and terminological simplicity

Throughout this study, the terms 'artificial agents' and 'conversational agents' are used as placeholder categories. In doing so, we knowingly pre-label these entities as agents, as artificial, and even as locally relevant properties of the setting. This does not imply that human participants themselves oriented to these features of the setting in such ways (or even that these *were* features of the setting for them). A more rigorous way of referring to these agents would be as 'intended-to-be conversational agents', i.e., artifacts designed to be treated as conversational agents. This will be the sense in which the terms 'conversational agent' and 'artificial agent' are used throughout this study. As [28, 60] have pointed out, whether interactions with LLM-based voice agents are in fact conversational should be treated as an empirical or analytic question rather than a premise.

---

7 Consequently, stemming from an EMCA perspective, this study is not concerned with the engineers and designers' 'intentions' that might serve as explanations of some ubiquitous features of voice agents or robots – nor with what features of these agents may simply result from unavoidable technical constraints, even if perfectly known to their creators. As Bittner [11] remarked, "[i]t seems reasonable that if one were to investigate the meaning and typical use of some tool, one would not want to be confined to what the toolmaker has in mind." The design history of artifacts (designers' constraints, goals, theoretical background, technical opportunities and current technological limits, etc.), what the toolmaker intended, thought, or did, is not required for an investigation of the situated use of technology, as long as this knowledge is not locally available to the participants themselves. Artificial agents are designed artifacts. Their conduct, shape, data streams [9], perceptual abilities (their metaphorical 'umwelt' [32, 102, 162]), conversational practices [12, 13], degree of transparency [88, 127], etc., are the result of complex social activities produced within material and technical settings. Hence, these agents' properties and behaviour – those encountered by so-called 'end-users' in natural interactions – are silent witnesses to fleeting prior design practices by industry professionals [125, 130]. Yet, the processes leading up to an artificial agent's 'final' form and behaviour are, overwhelmingly, not available as a resource for participants immersed in the immediacy of situated interaction.

8 As Schegloff [140] remarks, any pre-labelled exogenous structure (be it social or material) is "confirmed, reproduced, modulated, neutralized or incrementally transformed in that actual conduct to which it must finally be referred".

In other words, and inevitably, our analysis does not begin with a purely inductive stance – that is, with an "unmotivated" look at the data [134] prior to specifying the categories that are locally relevant for the participants. Any research in HCI or HRI necessarily involves an initial 'motivated' look that constructs its topic as part of that domain. As preliminary "analytic categories" [102], terms like 'agent' or 'robot' reflect an analytical commitment made by the researcher to engage with a particular scientific field. For example, the very notion of 'human–robot interaction' presupposes the local relevance of the categories 'robot' and 'human', as well as the existence of an 'interaction' between these pre-labelled entities. As an initial step, these pre-existing concerns suppose directing the analytic lens *toward* certain properties of the setting that are relevant for the researcher (i.e., robots or voice agents), before examining if and how these properties emerge as relevant for participants in situ[9].

### 3.3 Key concepts in this work

Human participants' practices are described using several concepts that have varied roots in ethnomethodology and conversation analysis. These concepts are described below.

1. *Omnirelevance of speech.* By omnirelevance of speech, we refer both to a technical feature of conversational agents – which triggers new turns every time a sound is identified as speech – and to a practical problem for the human participants. In the following excerpts, an orientation to the omnirelevance of speech was thus witnessable both in the contributions produced by the artificial agent *and* in the co-present humans' construction of their turns at talk. Speech was treated as omnirelevant for those human participants only insofar as they displayed a systematic concern that their spoken contributions may interrupt the voice agent's ongoing talk, or that speech not designed as interruptive or addressed to the conversational agent may nonetheless trigger a response.
    - For the conversational agent, speech was treated as omnirelevant in that it was systematically responded to whenever it was detected. From a technical point of view, this 'omnirelevance of speech' relates to a broader pattern of conduct observed in some voice agents or conversational robots: i.e., to respond to any input recognized as speech, even when this speech "pass[es] the opportunity" to talk [142] by being designed as response tokens [35] (or 'backchannels') such as "uh huh" [139]. Such speaking turns are responded to independently of whether they select a specific next speaker (through address terms [138] or through gaze orientation [5]) and of the sequential context in which they are uttered – e.g., as part of a storytelling [148] in which another participant intervening might be heard as interruptive.
    - For human participants, speech was treated as omnirelevant in that these patterns of behaviour from the robot became a practical problem to manage. The hearability of their voice to the conversational agent is a feature of their interaction that human co-interactants managed through their observable and hearable conduct, as any speech produced near the conversational agent may trigger a new unwarranted turn from this agent and disrupt the broader ongoing activity. Speech became omnirelevant for humans insofar as it was so for the agent. Thus, the orientation of human participants to an 'omnirelevance of speech' is a temporally bounded analytic result. It stems from a granular examination showing that, whether from the outset of the interaction or later, all participants' turns in the recorded interaction were demonstrably

9 Following the same logic, to remain entirely analytically rigorous, references to 'human voice' should not treat this property of a setting as an ontological feature inherent to any sound uttered by a human being. On the contrary, from a strict EMCA perspective, 'human voice' is a situated achievement. Consequently, sounds that are locally oriented to as 'human-voice' can, in principle, be produced either by biological humans or by machines.

designed so as not to interrupt the voice agent, or to trigger a response to speech that was not addressed to it.

2. *Aside sequences.* In response to the constant possibility of their voice being heard and responded to by the voice agent, human interactants built what we label 'aside sequences' that excluded the robot as a participant. These aside sequences partially differ from either the ordinary or technical concept of side-sequences [66, 73, 114, 131]:
   - Side sequences are brief, embedded exchanges that momentarily interrupt the main course of talk – for example, for clarification or repair – before returning to the primary activity [66]. These sequences are publicly available to all co-participants (including artificial agents) even when they are not *addressed* to these participants. For example (in contrast with Excerpt 2 below), during a game of cards around a table, 'giving private explanations to a beginner' can be produced while being audibly available to every seated guest, yet responded to only by members whom the explanation-giving turns are designed to address[10].
   - In contrast, aside sequences are not *designed to be recognizable* as 'private' for the other interactants; rather, they are *designed to be perceptually unavailable* to one or several participants (human or not). In the case of a robot or voice agent, these sequences meet the following criteria: by drawing on modalities that this particular artificial agent cannot detect[11] (e.g., certain types of gestures) and on its "sensor range" [47], aside sequences are strategically designed to exclude this agent as a participant. They are devoid of features that this specific robot or voice agent – whose abilities are known or identifiable by human participants over the course of the interaction – would register and respond to.

3. *"Voice User Interface speak"* [28].
   - We show that aside sequences are merely one ad hoc practice among others that facilitates the progressivity of the interaction with voice agents – by being designed so as not to elicit unwarranted responses from this agent. We illustrate this point by studying the incremental adaptation of a first-time user of voice agents. This participant initially brings established practices, drawn from everyday conversation, into her talk with the voice agent – before gradually abandoning them.

4. *Turn-taking systems and turn-taking models.*
   - In conversation analysis, *turn-taking systems* are the resources and practices through which participants construct turns and allocate speakership (i.e., who the next speaker is) [137]. Sacks et al. describe the turn-taking system for conversation in terms of a turn-constructional component (specifying what can constitute a complete turn, which, depending on what was said just prior, may be single words, or more expanded phrases or clauses), a turn-allocational component (distinguishing practices for other-selection and self-selection for the next turn), and a set of rules that participants orient to in producing, allocating and coordinating their turns [137]. What matters for the distinction with turn-taking models is the nature of these rules that govern turn-taking systems [16]. These rules are *normative rules* – used by participants as

10 Similar, though not entirely overlapping, publicly accessible explanation-giving frameworks are common in classroom settings, where "[t]he institutional role of teacher, therefore, involves being responsive not only to the student but also to the larger audience of students who can overhear what he has to say, even if the teacher addresses one student" [72]. However, both this configuration and the card-game example of private explanation discussed above fit the broad definition of 'side sequences' used in this paper.

11 These perceptual limits of the voice agent may either be discovered by participants during the interaction through the agent's patterns of conduct (see excerpt 3 for an example of this process of adaptation) or already understood beforehand.

resources during their situated activities – rather than computational rules executed deterministically [16]. Additionally, they are not statistical rules or regularities [22]. The turn-taking ruleset is not evidenced by establishing when participants take turns *most of the time* – i.e., not through a distributional analysis. Rather, the rules that compose a turn-taking system are normative resources that participants observably enforce or negotiate when establishing who will have the right or obligation to take the next turn.

- By contrast, we define *turn-taking models* as the statistical or deterministic mechanisms through which a conversational agent takes turns in *accordance* with the normative rules of the human turn-taking system. Indeed, specifying normative rules is not the same thing as constructing artificial agents whose conduct *aligns* with these rules. For example, any model trained on human conversational data that continuously predicts when to speak (e.g., [63, 157]) can operate in a strictly distributional way. It may estimate, from statistical regularities in the data, the probability that a speaker transition should occur at a given moment. Yet, in doing so, this model cannot be said to 'use' the normative rules participants orient to and that may explain this distribution of speaker transitions[12].

5. *Participation frameworks*.
   - A participation framework is the locally organized arrangement of who is taking part in an interaction, in what capacity, and with what access to the ongoing talk [44, 97, 156]. For the purposes of the following analyses, Hutchby [59] relevantly defines participation frameworks as "the range of ways that persons within perceptual range of an utterance are able to position themselves in relation to it; for example as addressed or not addressed, ratified or not ratified participant, and so on". In the excerpts studied in this work, a practical problem managed by human participants is that the voice agent will treat any utterance as addressed to it – or, at least, as requiring a response. Such an agent's conduct is not sensitive to participation frameworks in which a verbal contribution on its part is not warranted[13] (such as side sequences between two humans, or a long sequence of storytelling).

## 4 DATA COLLECTION AND ANALYTIC PROCESS

### 4.1 Data collection

The following analysis relies on video recordings of naturalistic interactions drawn from two corpora in which the spoken language is French:

- *Excerpt 1* is part of a corpus of 100 naturalistic interactions between visitors and the Pepper robot in a museum in 2022, in France – see Rudaz and Licoppe [131] for an extended analysis of this corpus. This corpus includes both dyadic and multi-party interactions. During this data collection, participants interacted with a conversational Pepper robot without having been given specific instructions as to what to say to the Pepper robot, or how to interact with it. All participants provided written informed consent before entering the room where the Pepper robot was installed, including consent to participate in the study, to have their interactions recorded, and to allow the use of their image. This data collection was approved by the Paris-Saclay Université Research Ethics Board. Participants in Excerpt 1 were not known to the authors of this paper.

12 That is, such models only rely on the "behavioral patterns" [79] that are the "outcome" [79] of a system of normative obligations.
13 See Umair et al. [163] about current challenges for conversational agents to appropriately respond within the framework(s) of a spoken interaction.

- *Excerpts 2 and 3* are drawn from a smaller video corpus of 11 naturalistic interactions that took place in 2025 between humans and OpenAI's multimodal "advanced voice mode"[14], in Brittany, France. These interactions feature human participants speaking with the voice agent on their smartphones while engaged in other encompassing activities, such as playing cards or reading the newspaper. These videos were recorded by the first author of this research, as part of an ethnographic study on older adults' initial or repeated use of voice agents. Excerpts 2 and 3 involve informants with whom a relationship had already been established by the first author. This author is involved as a participant in Excerpt 2, in which he is pseudonymized as Adrien. All participants in this corpus provided written informed consent regarding the recording of their personal data and the publication of images of themselves collected during this fieldwork. This data collection was conducted with the approval of the Research Ethics Committee of the University of Copenhagen.

### 4.2 Two types of artificial agents and their conversational capabilities

Two types of conversational agents are featured in the excerpts studied in this research:

1. *An exclusively rule-based 'social' robot (Excerpt 1)*. The Pepper robot used in this data collection relied on a simple rule-based chatbot designed to be 'conversational' and capable of responding to queries on a wide range of topics[15]. These data were collected before large language model-based conversational systems became widely available to the general public.
2. *A voice agent whose turn composition relies on generative AI (Excerpts 2 and 3).* ChatGPT's "advanced voice mode" generates its spoken contributions using a large language model, and was used in its multimodal version. Its responses were constructed from audible cues (captured by the smartphone's microphone) and visual cues (provided by the smartphone's camera feed).

However, despite the pronounced gap in technology and efficiency when it comes to *generating* their responses, both the artificial agents described above currently rely on silence-based turn-taking models when *positioning* their turns (see Section 2.4). These silence-based turn-taking models constitute the technical backdrop whose locally emergent properties (treated as resources, constraints or affordances) are scrutinized in the following excerpts. Thus, although the following sections examine broadly different dialog systems – with different response latencies and varying capabilities for recognizing human speech and generating appropriate contributions – the specific turn-taking mechanisms of these systems remain largely the same.

### 4.3 Data analysis and transcription

#### *4.3.1 Analytic Process*

The transcripts analyzed below originate from video recordings that were reviewed in detail across multiple data sessions. In the EMCA tradition, such sessions constitute a standard step of analysis [7, 48], bringing together researchers to closely inspect short segments of interaction. During these meetings, researchers debate [3] candidate formulations of the actions performed by the recorded interactants and the phenomena to which those actions are demonstrably responsive. Some aspects of this process have been characterized as a form of informal peer review [4], through which conversation analysts increase the robustness of their interpretations [10]. Each of the following video excerpts was examined during two data sessions involving all the authors of this paper, as well as during one or more additional sessions (depending on the excerpt)

---

14 For a detailed description of ChatGPT's "advanced voice mode", see https://help.openai.com/en/articles/8400625-voice-mode-faq. Accessed May 22, 2025.

15 See Rudaz [130], for a detailed description of the tools and standard practices through which Pepper was scripted until 2023

involving several experienced conversation analysts who were not involved in the authors' project. These repeated analyses allowed participants in the data sessions to reach a consensus on the main analytic claims presented in this paper.

#### *4.3.2Transcription conventions*

Participants' conduct was transcribed using Jeffersonian conventions for talk [68] and Mondada's multimodal conventions for embodied action [98]; further details are provided in the Appendix. Jeffersonian transcription is suited to the representation of subtle features of speech, including pauses, overlap, timing, and intonation. Mondada's conventions make it possible to annotate gaze, gesture, posture, and other embodied features in temporal relation to talk. Used together, these conventions allow the analyst to capture how spoken and embodied conduct are coordinated moment by moment [98, 101]. In the transcripts that follow, talk appears in bold, while embodied conduct is shown on the line(s) immediately underneath in regular font. Talk provides the reference timeline, and embodied conduct is synchronized to it through matching symbols (e.g., "&") placed vertically across the relevant lines. These symbols mark the onset or completion of embodied actions, such as shifts in gaze direction.

## 5 ANALYSIS

### 5.1 Producing 'aside sequences': Excluding the robot as a participant

In our first case of interaction with an artificial agent, two users (TOM and ANA) are concluding a semi-experimental trial interaction with a Pepper robot (ROB). TOM is standing in front of the robot and talking with it, while ANA is sitting at the round table and observing the scene (see Figure 1.1). As a whole, the sequence shown in the excerpt documents that the successful achievement of a smooth alignment, such as a mutual farewell sequence, requires careful work and precise timing on the part of the humans who ongoingly make sense of the robot's conduct, tentatively adjusting each next move or turn-at-talk in real time. As such, the accomplished semblance of harmony between the robot and its main user requires constant maintenance work on the part of the human participants. It is this work that we detail in the analyses below.

For purposes of clarity, the transcript and analysis of the video clip are divided into three shorter parts, with excerpts 1.1 and 1.2 explicating in detail the production of utterances directed at the robot and the production of aside sequences that are designed to remain inaudible to the robot. Finally, excerpt 1.3 shows what happens when TOM and ANA, after closing the interaction with ROB for the third time, relinquish their orientation to the omnirelevance of speech for the machine.

*5.1.1Excerpt 1.1: The first farewell*

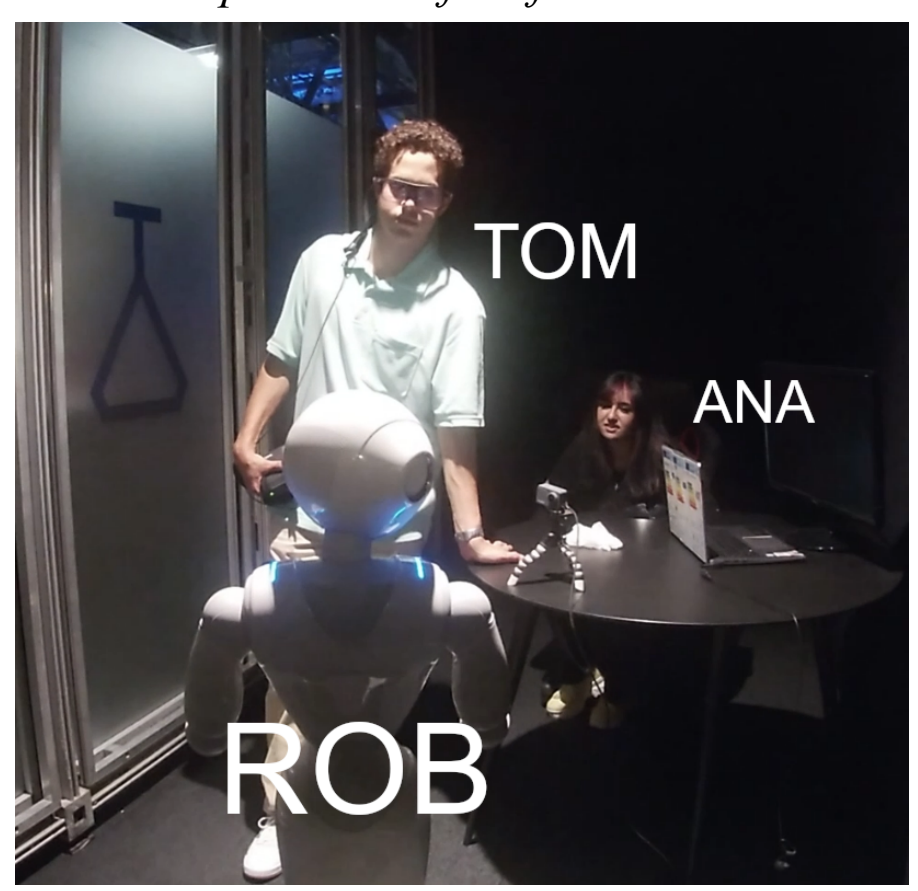

Fig.1.1

```
01 TOM    (alors/euh)
02        (0.5)
03 TOM    e%st-ce que [tu as beso]in >de tra<vai°ller°
          do you need to work
   rob     %screen shows '?'
04 ROB                [ comment ?]
                        what?
05        (1.3)
06 ROB    %#non (.) pas de tout
          no        not at all
          %screen shows 'Est-ce que tu as besoin de travailler'
   fig     #fig.1.2
```

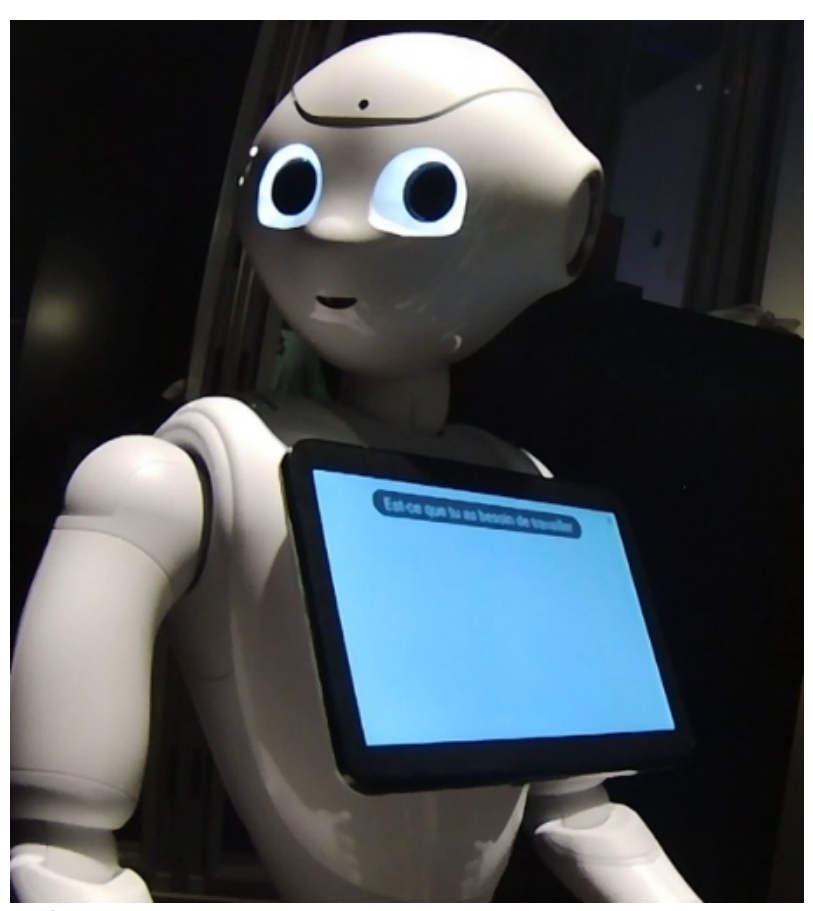

Fig.1.2

```
07        (0.7)&(0.2)
   tom         &nods
08 TOM    °d’ac&cord°
           alright
```

```
                &gaze twd ANA
09 ANA    (              )
10 TOM    °(tu as) ta question ?°
           (you have) your question?
11        (0.2)&(0.2)
   tom          &gaze twd ROB
12 TOM    .hh &au revoir# (.) bonne journée à t&oi↑#
          .hh goodbye         good day to you
   tom        &leans forward twd ROB          &straightens torso
   fig                  #fig.1.3                    #fig.1.4
```

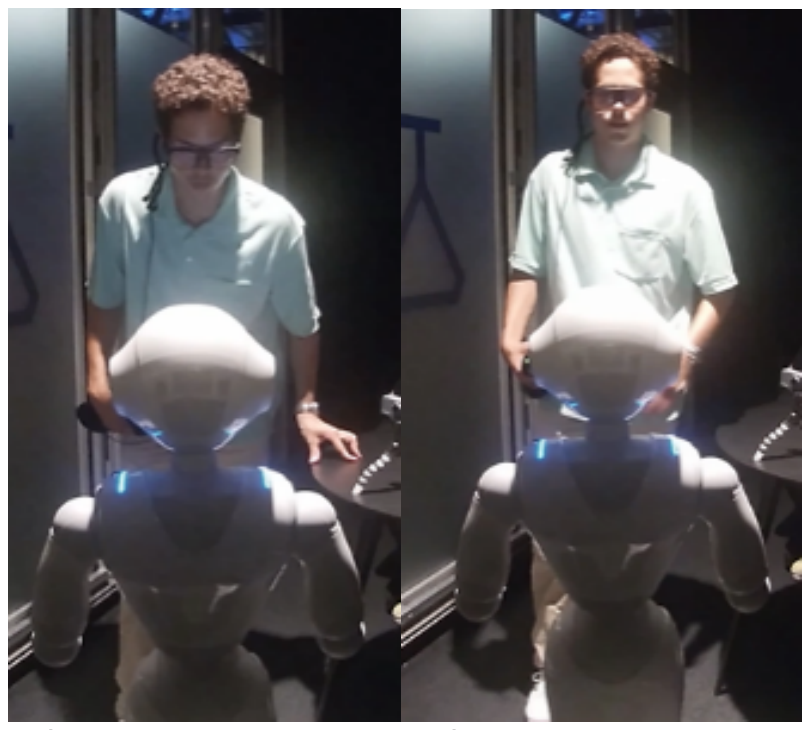

Fig.1.3        Fig.1.4

The first part of the transcript allows us to describe TOM and ANA's practices for interacting alternately with ROB and with one another. After a quiet and not clearly audible one-syllable utterance in l. 1, TOM orients to ROB and begins formulating the question starting with "Est-ce que" ("Do you", l. 3). Before he can finish the turn, the robot produces the repair initiator "comment" ("what", l. 4) – presumably still responding to TOM's first utterance in l. 1. The screen on ROB's chest (displaying the output of ROB's 'speech recognition algorithm', i.e., what ROB 'hears' from humans' speech) provides TOM with important resources for coordination. While ROB's screen displays a question mark[16], TOM completes his question slightly in haste. After 1.3 seconds of silence, the screen shows the transcription of TOM's question while ROB replies (l. 6), whereafter TOM acknowledges this response (l. 8). Then, following a short and softly spoken sequence of talking to ANA (l. 8 to l. 11), TOM visibly reorients to ROB again in l. 11 and produces a farewell greeting: "au revoir (.) bonne journée à toi" (l. 12).

Just how is it accomplished that some utterances are audibly done for ROB, and some can be clearly heard as designed to remain concealed from the machine? This directed character of utterances is manifest for the participants in the real time of the situation, as well as recognizable in retrospect for onlooking observers analysing the video recording. It is achieved through an ensemble of prosodic and embodied features of talk.

Utterances that are done specifically and evidently for the robot, such as the ones in l. 3 and l. 12, are produced in such a way as to be perceivable by the machine's sensors and intelligible for its processing algorithms. They are uttered at a higher amplitude [42] and at a slightly slower pace than surrounding speech, with marked prosody and clearer pronunciation. Sequentially, they are set off from the preceding talk by brief pauses (l. 2, l. 11). At the same time, the talk done for the robot is also produced in a bodily posture that displays a distinct orientation to it. TOM's gaze is directed at ROB, and his upper torso is leaning forward (see Figure 1. 3), thereby placing his vocal tract closer to the machine's

16 From a technical point of view, this question mark indicates an issue to recognize the previous utterance produced by the human.

microphones. Speaking this way, TOM displays an emergent competence as a ‘robot user’, orienting to the sensory capabilities and restrictions of the robot [113], and methodically reducing the occasioned need for repair. Concurrently, ANA displays this competence by remaining silent while TOM is talking with the robot.

By contrast, utterances manifestly done by TOM and ANA for each other, and specifically not to be perceived by the robot’s sensors, are notably lower in volume. In fact, they are produced so quietly that they are barely picked up by the recording devices placed in the room by the lead author of this article. This is visible in l. 8 to l. 10, which comprise an exemplary sequence that excludes ROB as a hearer, and the lack of response from ROB can be seen as proof of success. In addition, TOM’s gaze also shifts from the robot in the direction of ANA, and because of ANA’s location in the room, this shift also requires a slight body torque [144] towards TOM’s left. In short, such aside sequences between humans in the presence of a robot are built in a way that shows an orientation to the omnirelevant possibility of their voice being heard and responded to by the robot, and they prevent this from happening.

The following transcript shows how the interaction continues after TOM’s first farewell in l. 12.

*5.1.2Excerpt 1.2: The second farewell*

```
12 TOM    .hh au revoir (.) bonne journée à toi↑
13        (0.9)%(0.1)
   rob         %screen shows 'Au revoir bonne journée à toi'
14 ROB    tu dois part(ir) ?
          do you have to leave?
15 TOM    =oui, désolé, &il faut# que j'aille
           yes  sorry    I've got to go
   tom                  &LH thumb points over L shoulder
   fig                          #fig.1.5
16        (0.5)
```

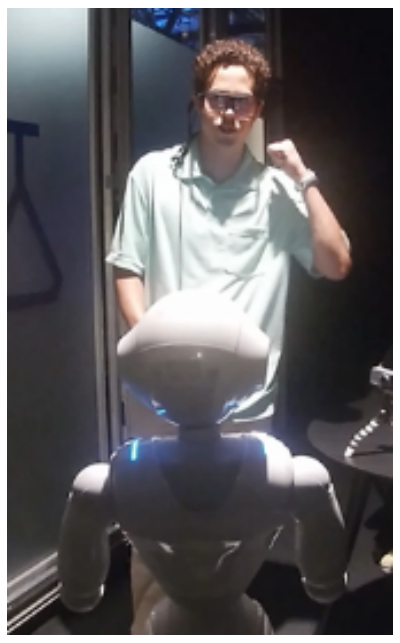

Fig.1.5

```
17 ANA.   °tu n'es pas très intelligent°
           you are not very smart
18 ROB    %entendu
          understood
   rob    %moves arms forward-->
          %screen shows 'Oui désolé il faut que j'aille'
19        (0.5)
20 TOM    pas de%soucis↑%#
          no worries
   rob        ->%LH+RH on hips%
   fig                    #fig.1.6
```

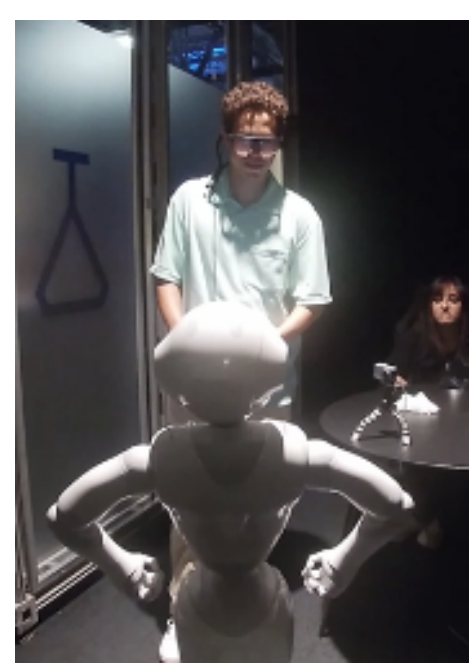
Fig.1.6

```
21          (0.2)%(0.4)
   rob          %raises LH-->
22 TOM       pas de soucis.
             no worries
23           (0.8)%#(0.8) & #(0.2)  #(0.5)
   rob         ->%LH fist stretched up
   tom                 &RH fist, open palm, fist
   fig            #fig.1.7 #fig.1.8#fig.1.9
```

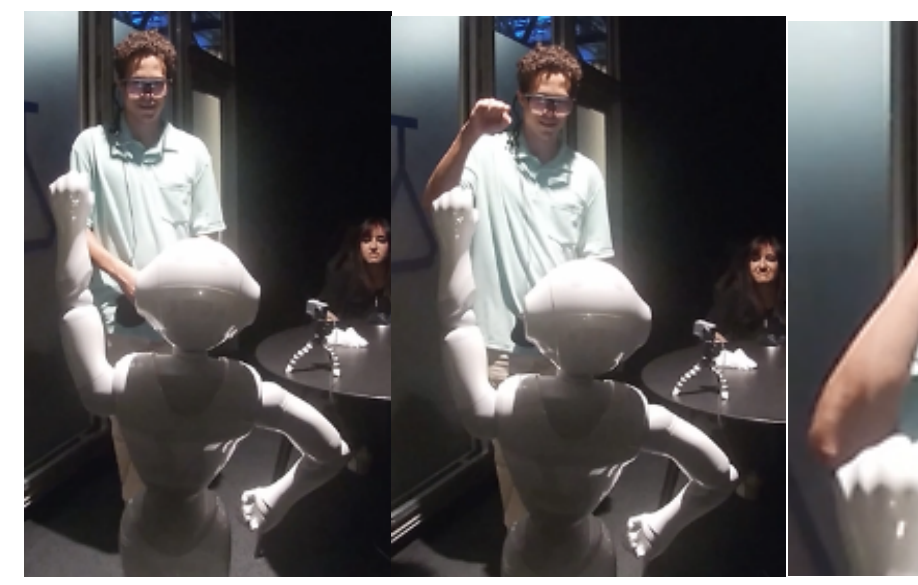
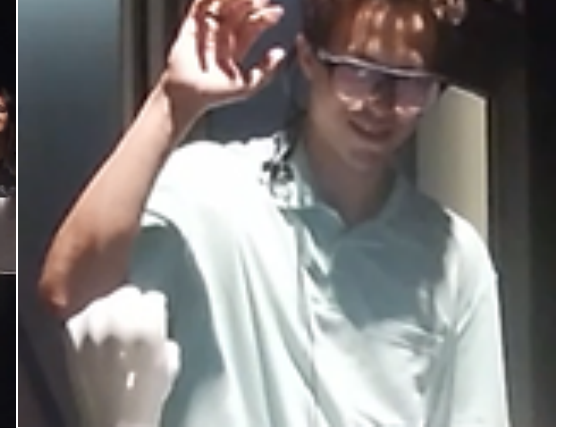
Fig.1.7 Fig.1.8 Fig.1.9

```
24 TOM      sa#lut&
            bye
   tom            &puts RH down
   fig        #fig.1.10
```

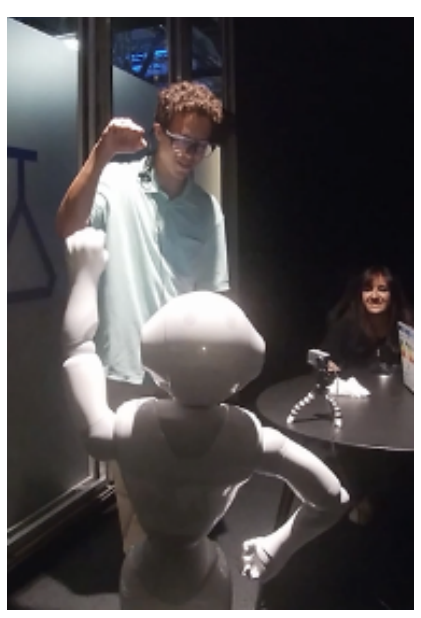
Fig.1.10

```
25          %(1.4)
```

```
   rob    %puts RH down
26 ROB    bonne journée
          have a good day
27 TOM    bonne &journée
          have a good day
   tom          &turns twd ANA, gaze at ANA
```

Although l. 12 in excerpt 1.1 is recognizable as a valediction, it is also a rather abrupt ending of the ongoing interaction that comes without any preliminary closing moves [147]. In that sense, ROB's response in l. 14 seems properly placed, inquiring about the possibly absent account for the closing: "tu dois part(ir)?" ("do you have to leave?"). TOM's next turn in l. 15, which includes an agreement and an apology, seems to affirm the adequacy of the robot's question. Nevertheless, ANA's critical assessment of the robot's 'intelligence' in l. 17 is strikingly different; interestingly, its prosody suggests that it is designed to be heard only by TOM but syntactically phrased as if it were addressed to ROB. Next, in l. 18, the robot produces an acknowledgment token, responding to TOM's explanation that he must go (l. 15).

At the same time, the robot starts to move its arms forward in l. 18. While TOM repeats "pas de soucis" ("no worries") twice in l. 20 and l. 22, ROB continues to perform a transformation of its bodily position, first putting hands on hips (Figure 1.6), and then stretching its left arm upwards with a closed fist (Figure 1.7). For a moment, TOM inspects ROB that is not moving anymore, and then hesitantly puts up his right hand in a position similar to the robot (Figure 1.8), briefly opening his palm (Figure 1.9), and then shaping it again into a fist. The peak of this continuous sense-making moment, and of TOM's embodied work, is achieved in l. 24 (Figure 1.10), when both ROB and TOM have their fists up in the air in an improvised, locally developed greeting gesture. The mutual alignment and sense of this position is affirmed by TOM's "salut" (l. 24), accompanied by a short thrust of his fist towards ROB's (without touching it). Both TOM and ROB then put their hands down, and the interaction seems to be properly closed as they wish each other a nice day (l. 26–27). While no full aside sequence occurs in this excerpt, the excerpt demonstrates how a turn-at-talk may be produced in a way that excludes ROB from the interaction.

The following transcript shows how the event continues after TOM's second farewell in l. 27.

### *5.1.3Excerpt 1.3: The third farewell*

```
27 TOM    bonne journée
          have a good day
28 ANA    °bonne journée, au non-rev[o  i  r°]
           have a good day see you never
29 ROB                              %[tu  dois] partir?
                                      do you have to leave?
   rob                              %screen shows 'Bonne journée'
30 TOM    &OUI (.) OUI (.) oui je:: je dois y aller
          YES      YES     yes I:: I have to go
   tom    &turns twd ROB, slight step twd ROB
31        (0.4)
32 TOM    je dois partir
          I have to leave
33        %(0.1)
   rob    %screen shows 'Oui oui oui je je peux y aller'
34 ROB    d'a%ccord
          alright
   rob       %goes through the same series of arm movements as in line 21-->
35        (1.2)
3 TOM    °pas de soucis° &(.) salut
```

```
          no worries (.) bye
   tom                   &RH quick wave, turns twd ANA
37 ANA    hhheh
38 TOM    (il y a    )
           there is
39 ROB     %%bonjour.#
             hello.
   rob       %screen shows 'salut'
   rob    ->%
   fig               #fig.1.11
```

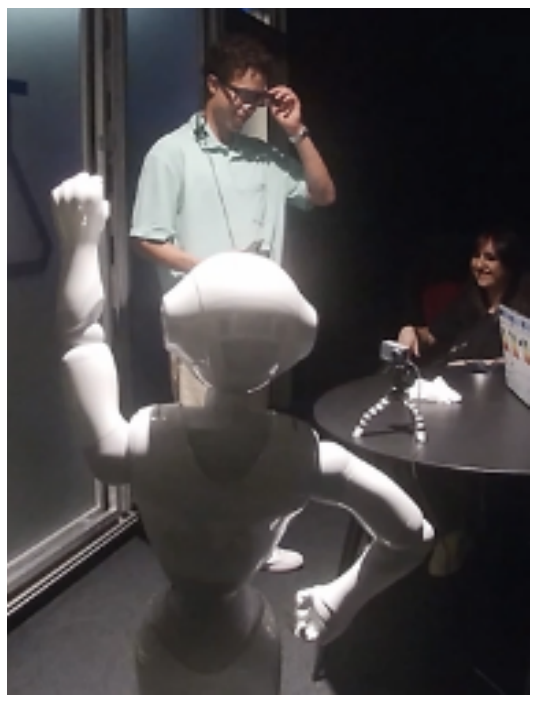

Fig.1.11

```
40 ANA    .hhh hhhehhh (.)$ah la la
   ana                    $gets up to leave
41 TOM   (tu peux commencer)
          you can start
42 ANA    [j'ai exacte]ment la même chose chez moi
            I have exactly the same thing at home
43 ROB   %[ comment ? ]
             what?
          %screen shows '?'
((ROB waves. Meanwhile, TOM and ANA do not respond to nor look at the robot anymore;
they talk to each other and leave the room))
```

Following the previous exchange of greetings, in l. 28, ANA contributes a sarcastic comment which overlaps with ROB's reopening of the seemingly closed conversational exchanges. In l. 29, the robot repeats the question it asked previously in l. 14 – "tu dois partir"? ("do you have to leave?"). Thereafter, TOM turns back to the robot and repeatedly agrees (l. 30), speaking loudly. After a short silence, in l. 32, he repairs his assertion by replacing "aller" ("to go") with "partir" ("to leave"), using the same verb as ROB, in an apparent attempt to make his talk more intelligible for the machine. After the robot minimally responds in l. 34, TOM offers a final farewell greeting (with a brief waving gesture) in l. 36. This is effectively the end of the interaction with ROB for TOM and ANA, as they start leaving the room.

This conspicuous reorientation is crucial with respect to the focus of our analysis, as both human participants relinquish the previously maintained participation framework characterized by alternation between turns produced covertly among themselves and those done hearably for ROB. After l. 36, the robot continues to be ignored as it once again goes through the same series of hand movements culminating in a left arm stretched up with a fist. In l. 39, in the position captured in Figure 1.11, it seems to commence a new interaction 'cycle' in response to the farewell ("bonjour"), then a repair initiation

(l. 43), and a waving gesture (not transcribed). However, none of this conduct of the robot is responded to by TOM and ANA anymore, in what seems to be an interactional equivalent to turning the machine off. They continue talking to each other in a way that no longer orients to the omnirelevance of speech for ROB, and simply disregard its attempts at interacting with them and responding to their voices.

In sum, an orientation to the omnirelevance of hearable speech as a designed feature of the robot transpires throughout the three parts of the analysed video clip. The feature is oriented to by TOM and ANA in the way they speak to each other, and specifically *for* one another, as noted above. Given that any instance of a human voice can be picked up and responded to by ROB, and that the situation is simultaneously a 'technology trial' and a 'fun shared experience', the local issue for TOM and ANA is to devise spatially differentiated methods of talking recognizably for the robot or for one another. On the part of ROB, the omnirelevance of speech seems to be behind the misplaced repair initiators in l. 4 and l. 43, both of which are treated as irrelevant and 'sequentially deleted'. For the robot, speech is 'omni-relevant' in the sense that it cannot be disregarded, and each turn is treated as occasioning a response. Concurrently, it is indeed only and precisely 'speech' that ROB is oriented to, unable to maintain common interactional history or any locally relevant individual personalities, beyond the machine-schematic role of 'the user' as a series of consecutive turns at talk. Once the interaction is evidently and recognizably closed – and re-closed – for TOM and ANA, an occurrence of human voice delivering intelligible content ("salut") seems to trigger an opening of a new interaction sequence for the machine, regardless of the fact that the participants being greeted are identical to the ones who have just said their farewells.

### 5.2 'Staying under the radar': Orienting to the voice agent's speech detection threshold

Our next series of excerpts shows an interaction with a voice agent (VOA) that is instructing a beginner player, Cédric (CED), how to play a card game. CED is interacting with VOA, whose verbal contributions are mediated by a smartphone that CED holds in his hand. There are two additional participants around the table: Guillaume (GUI), and Adrien (ADR), who are more knowledgeable about the game. They overhear VOA's instructions to CED, aligning with the activity but not always affiliating with the instructions given. The voice agent featured in this interaction is ChatGPT's "advanced voice mode". As mentioned in Section 4.1, ADR is the first author of this paper and participates in the game both as a player and as an ethnographer collecting data for a study of elderly individuals' use of voice agents. This research objective had been explained beforehand to the other players.

All through the excerpt, the participants treat CED as the VOA's sole interlocutor in a one-on-one participation framework overheard by the other players. Holding the smartphone that receives both sound and image, CED recurrently asks VOA for the next instructions on how to play, and he also verbally responds to them. However, as the excerpt shows, CED and the other two human participants simultaneously manage a distinct participation framework where the VOA is, similarly to excerpt 1, effectively 'shielded off' from the interaction by minimizing the sound of their voices. Just as with our first case, we have chosen to divide this excerpt into three separate parts. When the first part begins, VOA finishes explaining what a 'joker' is.

*5.2.1Excerpt 2.1*

```
01 VOA    elle peut remplacer n'importe quelle autre carte
          it can replace any other card
02        pour t'aider à former une combinaison gagnante.
          to help you form a winning combination
03        (0.4)
04 CED    oké. .hh alors. (.) maintenant explique-moi là.=
          okay .hh so         now explain to me
```

**05**         **=là j'suis avec euh: un- un ami:, (.) explique-moi (pou)=**
*I am with a a friend explain to me (to)*
**06**         **=qu'est-ce qu'on doit faire pour- pour jouer.**
*what one should do to to play*
**07**         **#(2.1)**
fig    #fig.2.1

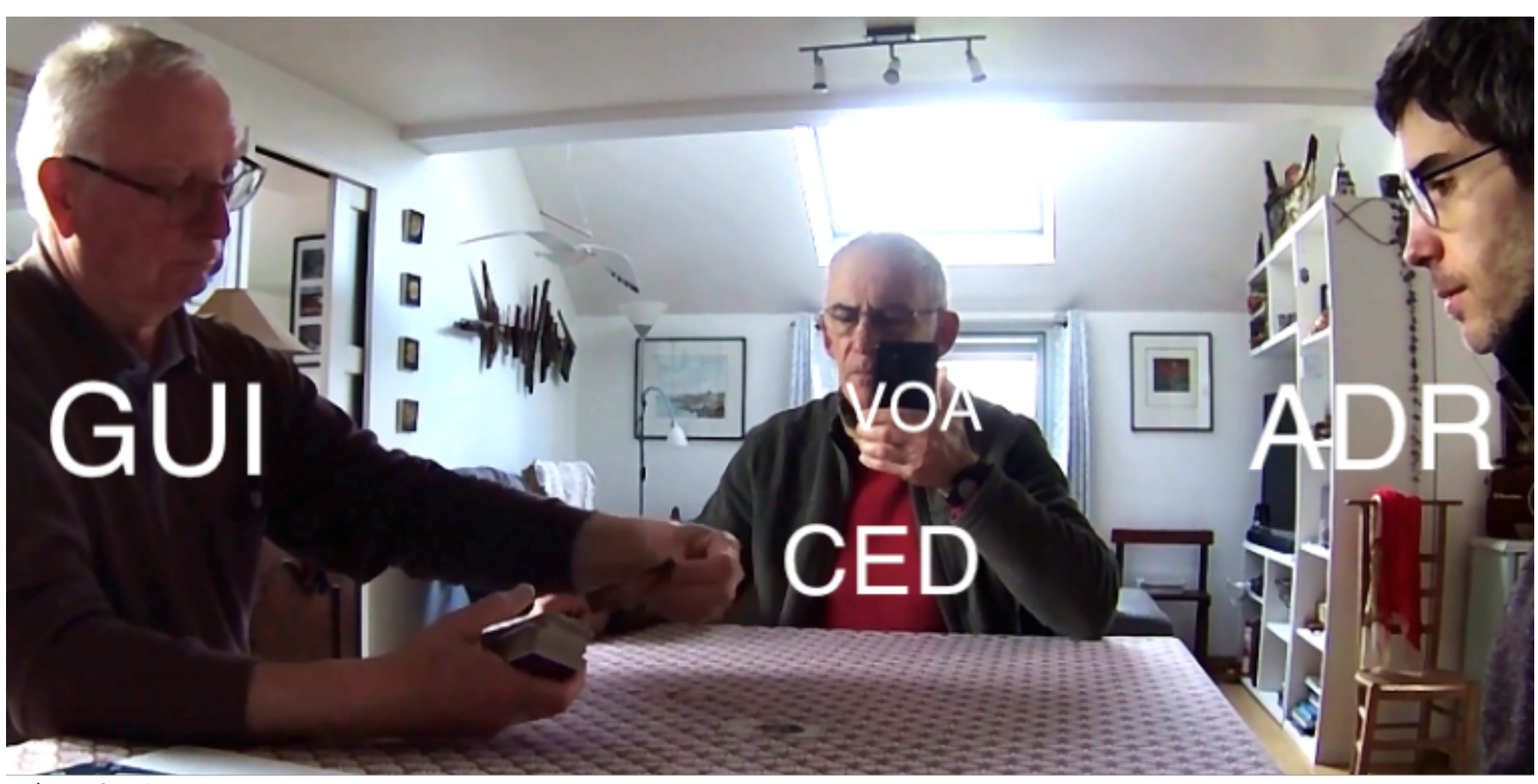

Fig.2.1

**08 VOA**     **d'accord. (0.5) pour commencer,**
*ok to begin*
**09**         **(.) chaque joueur reçoit un certain nombre de ca+rtes.**
*every player receives a certain number of cards*
gui                                                      +dist cards-->
**10**         **(comme on va) [en ]suite,**
*(as one will) then*
**11 CED**     **[combien],**
*how many*
**12 VOA**     **vous essayez+de former+des combinaisons,**
*you try to form combinations*
gui            -->+.........+three fing-->
**13 VOA**     **#comme des+suites,+ (.) ou des groupes de cartes de même valeur.**
*like straights or groups of cards of the same value*
gui            -->+,,,,,,,+
fig    #fig.2.2

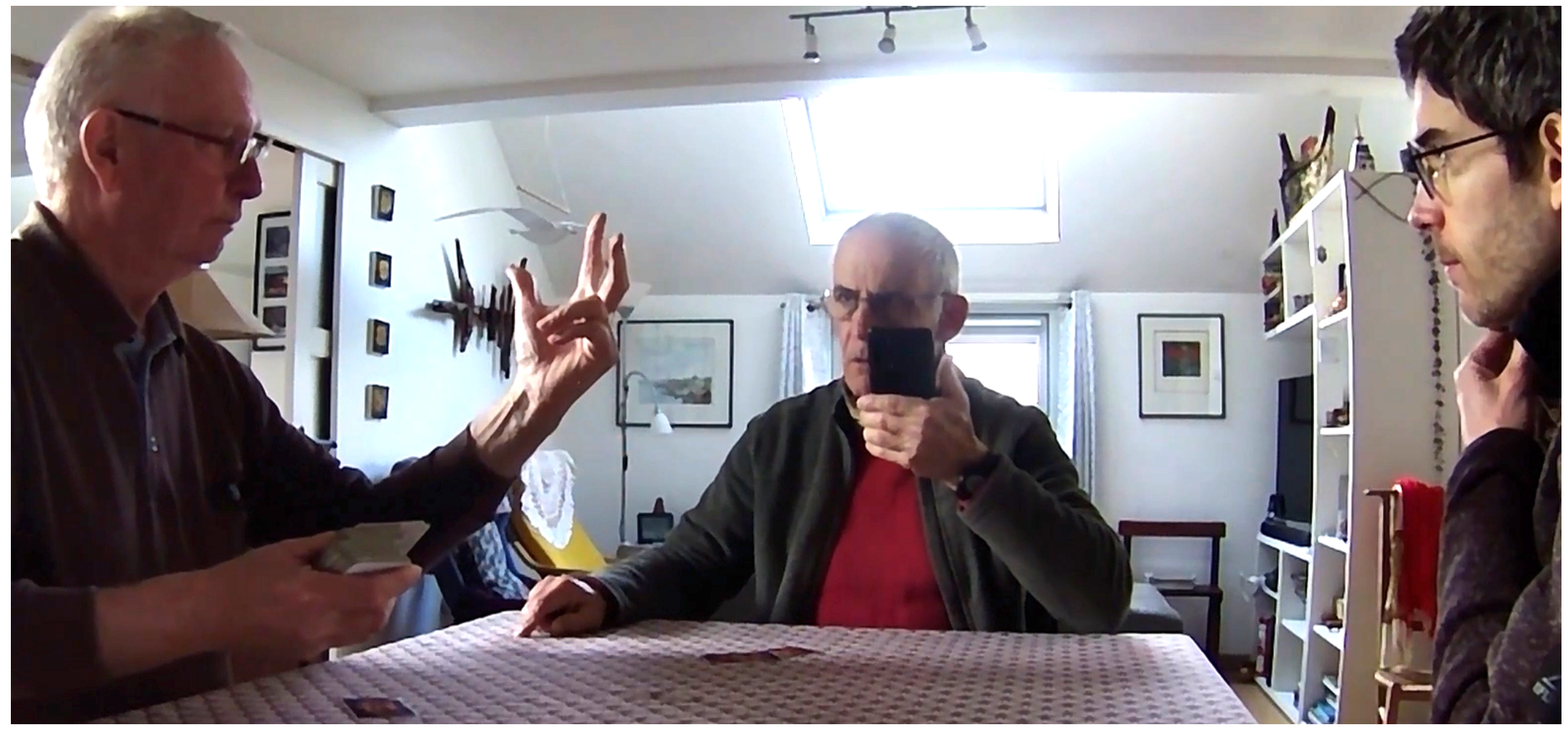

Fig.2.2

At the end of VOA's explanation of what a joker is, CED moves on to ask VOA how he and the friend he is with should actually play the game (l. 4-6). When VOA begins its reply by stating that everyone should get a certain number of cards (l. 9), GUI starts to distribute cards on the table. While GUI is distributing the cards to each participant in a clockwise fashion, CED, the beginner player, requests that VOA should be more precise about the number of cards (l. 11). He gets no immediate answer as VOA continues to instruct on other aspects of the game (l. 10-13). Instead, it is GUI who responds – momentarily pausing the card distribution – using a gesture with three fingers in the air in front of CED as well as forming the word "trois" (three) with his mouth (l. 11-12, Figure 2.2). He thus responds to the question, despite it not being addressed to him, orienting to a preference for a response to be produced [33, 152]. However, as he does this through visual means – using gaze and gesture – his response is not readily perceivable by VOA, and in this way does not claim the turn relative to VOA. This first part of our excerpt, where GUI is mouthing and gesturing his answer to CED's question, thus demonstrates, in addition to the relevance of the production of an answer to CED's question, GUI's orientation to VOA's capacity for capturing the sound of his voice. By designing his answer silently, he manages to stay out of the interaction between CED and VOA.

*5.2.2Excerpt 2.2*

```
14 CED     (y a com[bien d'ke-) >attend 'tend] 'tend<, doucement,
           (how many xx)         wait     wait   wait   take it easy
15 VOA             [(tel que:::)             ]
                    (as for instance)
16 CED     combien de cartes à: chaque: joueur.
           how many cards for each player
17         (2.5)
18 VOA     d'accord.
           ok
19         (0.3)
20 VOA     en général, (.) chaque joueur commence avec dix cartes.
           generally       every player begins with ten cards
21         +&(0.3)&(0.2)+@vous=
                          you
```

```
   gui    +............+gaze CED-->
   ced     &.....&gaze GUI, raises chin-->
   adr               @...-->
22 VOA    =pou#@vez+   [ajust   +      ]er#selon vos     +pré@fé&rences,+
           can         adjust             as you wish
23 ADR             +   [°dix car+tes¿°]
                        ten cards
   ced                                                       -->&
   gui          -->+............+gaze ADR, three fingers-+,,,,,,,,,,,,,,+
   adr      -->@gaze GUI-------------------------------------@
   fig          #fig.2.3                    #fig.2.4
```

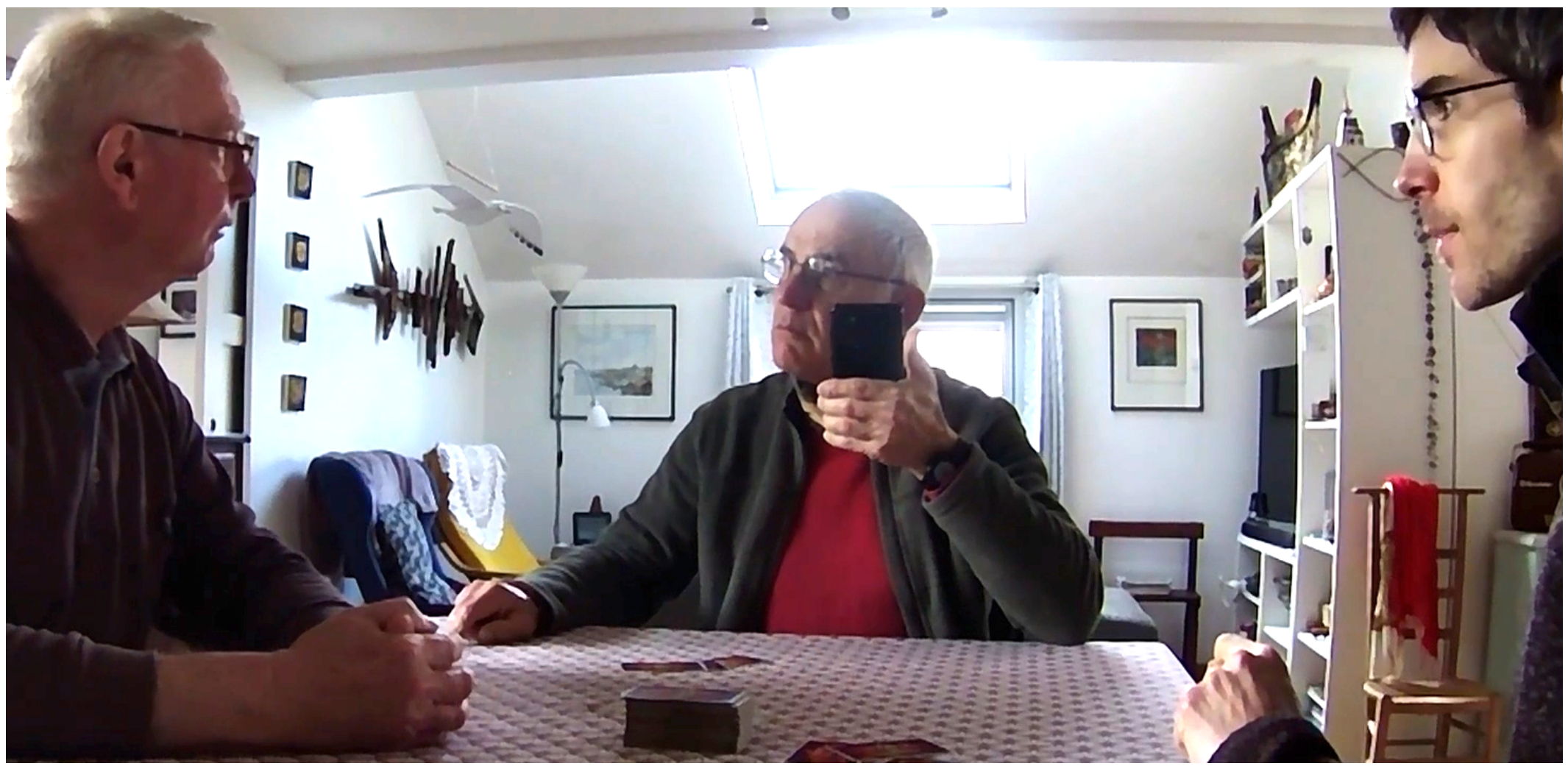

Fig.2.3

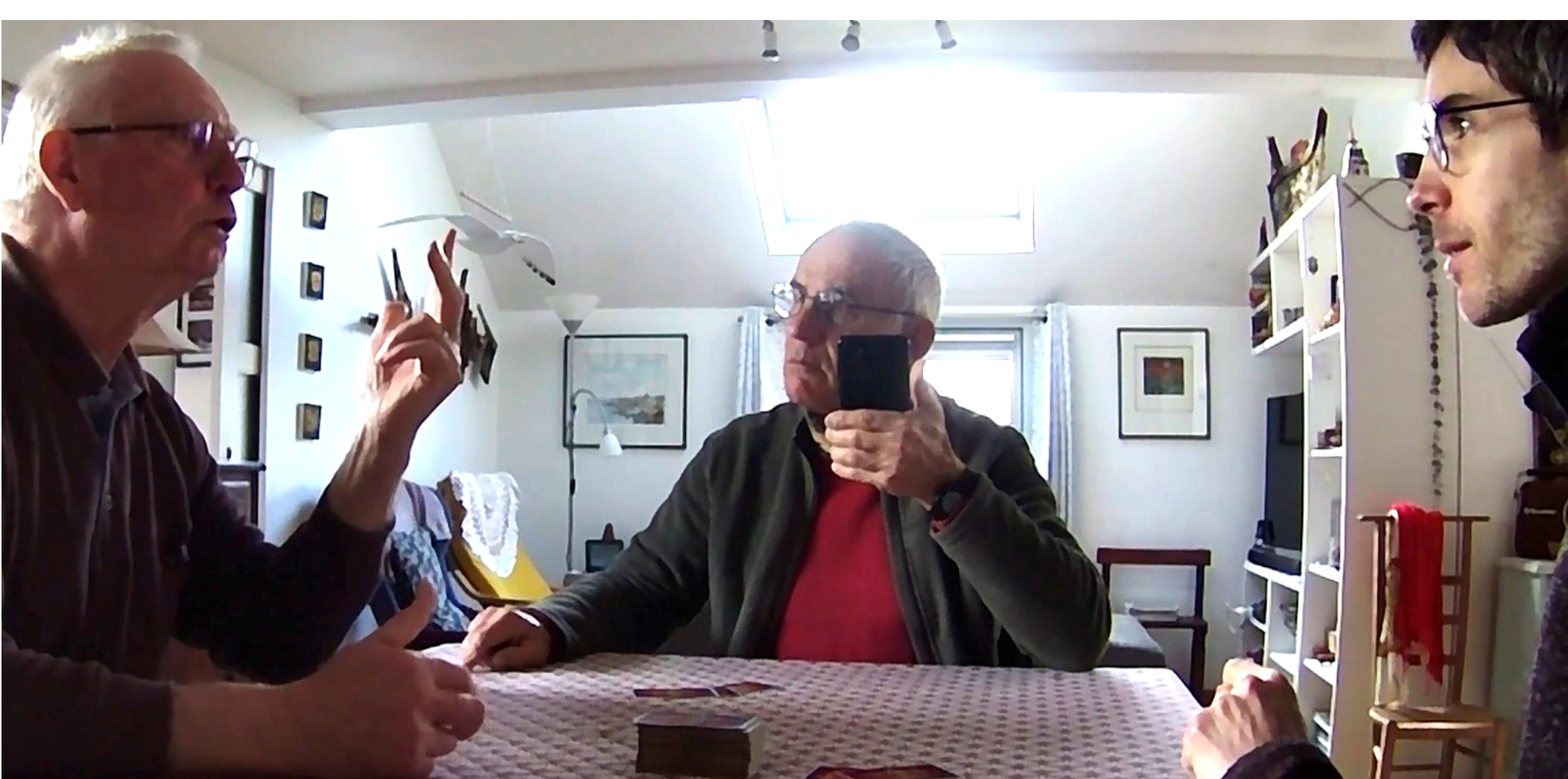

Fig.2.4

```
24.    (.) mais dix cartes, est u+ne bonne base pour débuter.
           but ten cards   is a good way to start
   gui                         +shakes head-->
```

Although GUI has already provided an answer to CED's question about the number of cards to be used (l. 12-13 above), CED nevertheless pursues an answer also from VOA, interrupting its continued instruction to repeat his request in a more explicit way (l. 14-16). By asking about the required number of cards in a way that is sequentially tied to VOA's prior talk, but with no reference to the response he has already received from GUI, CED treats VOA as lacking access to the talk occurring between human participants. In return, VOA first produces a receipt token ("d'accord"; ok, l. 18) and then moves on to explain that one generally starts with ten cards each (l. 20). This number stands in sharp contrast with the "three" cards that GUI just mentioned; VOA's explanation does not index the previous assertions produced by GUI regarding the correct number of cards. This conduct aligns with CED's treatment of VOA as having no perceptual access to the verbal exchanges produced between the human participants. As soon as VOA produces "dix cartes" (ten cards) (l. 20), the human participants display a critical stance towards this instruction. GUI turns his gaze to CED, who returns the gaze in silence and raises his chin (l. 20, Figure 2.3), while ADR questions the information given by addressing a softly produced request for confirmation "°dix cartes¿°" (l. 23) to GUI. Turning away from CED, GUI shifts his gaze towards ADR as he redoes the three-finger gesture, again silently mouthing the word "trois" (l. 22-23, Figure 2.4) and shaking his head (l. 24). In doing so, he affiliates with the critical stance hearable in ADR's request for confirmation and insists on three as the adequate number of cards.

This part of our excerpt thus offers further demonstration of how the human participants build two distinct participation frameworks, where only one includes VOA. In addressing the smartphone without reference to the ongoing or prior 'human-human' interaction, CED treats VOA's contributions as produced in complete isolation from the other participants' co-occurring verbal and gestural conduct. Yet, each time CED finishes addressing VOA, the human participants use gaze, iconic gestures, facial expressions, head movements, and mouthings (i.e., mainly visually based communicative resources), as well as softly produced talk, to display their stance toward VOA's response. In doing so, they work to *produce* VOA's exclusion from what emerges as their private participation framework.

Excerpt 2.3 is the last part of the excerpt. As it begins (l. 25-27), CED refers to the wish by his friend to play with three cards instead.

*5.2.3Excerpt 2.3*

```
25 CED    d'accord (mais) voilà il préfèr+eu[::         ]
          ok       (but)  PART   he prefers
26 GUI                                       [°(à) trois°]
                                               (with) three
   gui                                 -->+
27 CED    qu'on fasse à trois. (.) trois cartes.
          that we do it with three three cards
28        (0.2)+(0.7)
   gui         +gze at/above ADR-->
29 ADR    @(           x+         xx)+@
   adr    @gze GUI quick head tilt---@
   gui                 -->+three nods-+
30 VOA    d'accord. (0.3) trois carteus=
          ok              three cards
31.       =chac+un, (.) c'est+par@fait [pour   ] commencer.
           each          it's perfect    to start with
```

```
32 ADR                                 [°°heh°°]
   gui         +tilts head fw+
   adr                           @smiles-->
33 CED    (.) (°alors+ou(h)i¿°)
              (so yes)
   gui               +nods once-->
34 VOA    mainteunant, (.) cha+cun essaye de@former=
          now              each tries   to form
   adr                                   -->@
   gui                     -->+
35        =des combinaisons avec ses cartes. (0.5) le but
           combinations with their cards           the goal
36        (.) est de se débarrasser de ses cartes en formant des suites,
              is  to get rid of        one's cards by forming straights
37        (.) ou des groupes.
              or groups
38.       +(.)+
   gui    +...+
39 GUI    #+°°voilà°°         +
          that's it
   gui     +gze ADR, beat gest+
   fig    #fig.2.5
```

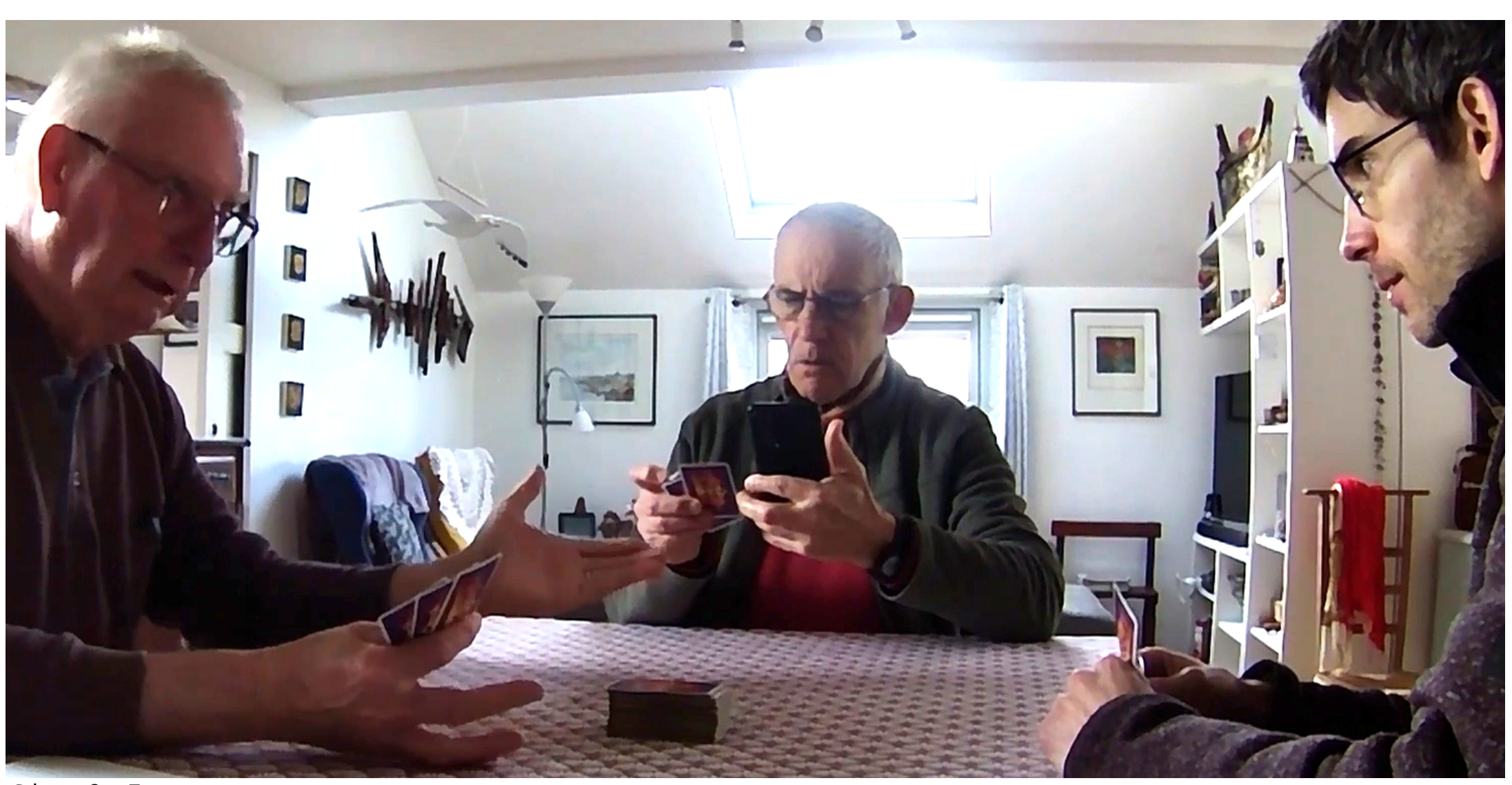

fig.2.5

```
40        +(.)
   gui    +,,,-->
41 ADR    °(d'a+cc[ord)° ]
            ok
42 CED            [°d'acc]ord,°
                    ok
   gui      -->+
43        (0.8)
44 CED    alors- (1.4) est-ce que- est-ce que tu vois mon jeu,
```

so do do you see my cards

After the short pause that follows CED's informing VOA that they prefer to play with three cards instead of ten, GUI and ADR nod in agreement, while ADR delivers a turn in a very soft voice (l. 29). VOA affiliates with CED's stated preference by uttering "d'accord. (0.3) trois carteus chacun," ("ok three cards each", l. 30-31). It immediately continues with a claim that this number is perfect to start with, with no reference to its own, previously very different suggestion. ADR's subsequent conduct is responsive to this contradiction: he smiles and issues a soft and minimized laughter token (l. 32). The remainder of the excerpt, where VOA continues the instruction, displays further audibly minimized actions by ADR and CED in doing agreement (a whispered "voilà", together with a beat gesture, l. 39, Figure 2.5) and in displaying understanding (a whispered "d'accord", l. 39 and 41). The excerpt ends as CED continues his direct interaction with VOA, again with a clearly audible voice. Overall, it offers additional evidence of how the three human participants accomplish and maintain a private participation framework, designed to be imperceptible to VOA. Listening to CED's interaction with VOA as overhearers, GUI and ADR make sense of and comment on the situation mainly through visual and/or soft-spoken contributions: 'giving private explanations' is here accomplished mainly through mouthings (as GUI does l. 22 – 23) and gestures. As indicated in Section 3.3, these practices constitute aside sequences in that they are *designed to be audibly unavailable* to the artificial agent. These aside sequences were not built to be accountably '*private'* to VOA (i.e., perceivable and recognizable by VOA as private exchanges between human participants) but rather to be *unavailable* to VOA. They were often whispered, mouthed, or gestured – rendered perceptually inaccessible to the artificial agent, yet still recognizable by human co-participants. Unlike typical side sequences, aside sequences were constructed to be 'off the record', i.e., not to be overheard or responded to by VOA. They allowed participants to negotiate meanings or alignments without triggering the agent's turn-taking machinery.

In total, the three parts of this excerpt have repeatedly shown how participants orient, in their embodied behaviour, to what may be thought of as an assumed 'amplitude threshold', below which it is possible to interact without the agent perceiving it. In doing so, participants *demonstrate a working understanding of the machine's sensorial abilities.* Keeping their voices very soft and otherwise using gesture, gaze, and head movements, they manage to establish a participation framework that excludes the agent and where they display their stances to the instructions that VOA produces. That these visual and soft-spoken contributions are also available to CED makes him a party to both frameworks.

### 5.3 Learning "Voice User Interface speak": Treating one's response tokens as 'slips of the tongue'

The practices described in excerpt 2 – produced in a multiparty interaction involving humans displaying familiarity with recent voice agents – can be insightfully contrasted with the methods initially enacted by a complete novice user of voice agents, in a dyadic interaction. Such is the case in the next excerpt: it features Emma's (EMM) first-ever interaction with a voice agent. Such situations can be heuristic in that they engender a clash between routinized conversational practices developed over one's life – from and for human interaction – and 'what works' [90] when speaking with a voice agent. Put differently, in the following excerpt, EMM's initial expectancies were not shaped by interactions with voice agents; they are not the result of prior trial-and-error with these devices.

The following episode takes place as EMM uses a smartphone equipped with a voice agent to read the newspaper (Figure 3.1). The system EMM uses is the same as the one featured in the previous excerpt, namely OpenAI's multimodal "advanced voice mode". As mentioned, this system can draw on input from the smartphone's camera to construct its verbal responses. Relevantly for the activity visible in this excerpt, EMM cannot read without specialized assistive devices. She is a visually impaired person with macular degeneration and has no vision in the center of her visual field.

*5.3.1Excerpt 3.1*

**01 GPT ΔΔI see the# first paragraph starts with "chers [citoyens]"**
*"dear citizens"*
emm >>Δgazes towards camera feedback on smartphone screen-->
emm >>Δorients the smartphone's camera towards newspaper>>
fig #fig.3.1

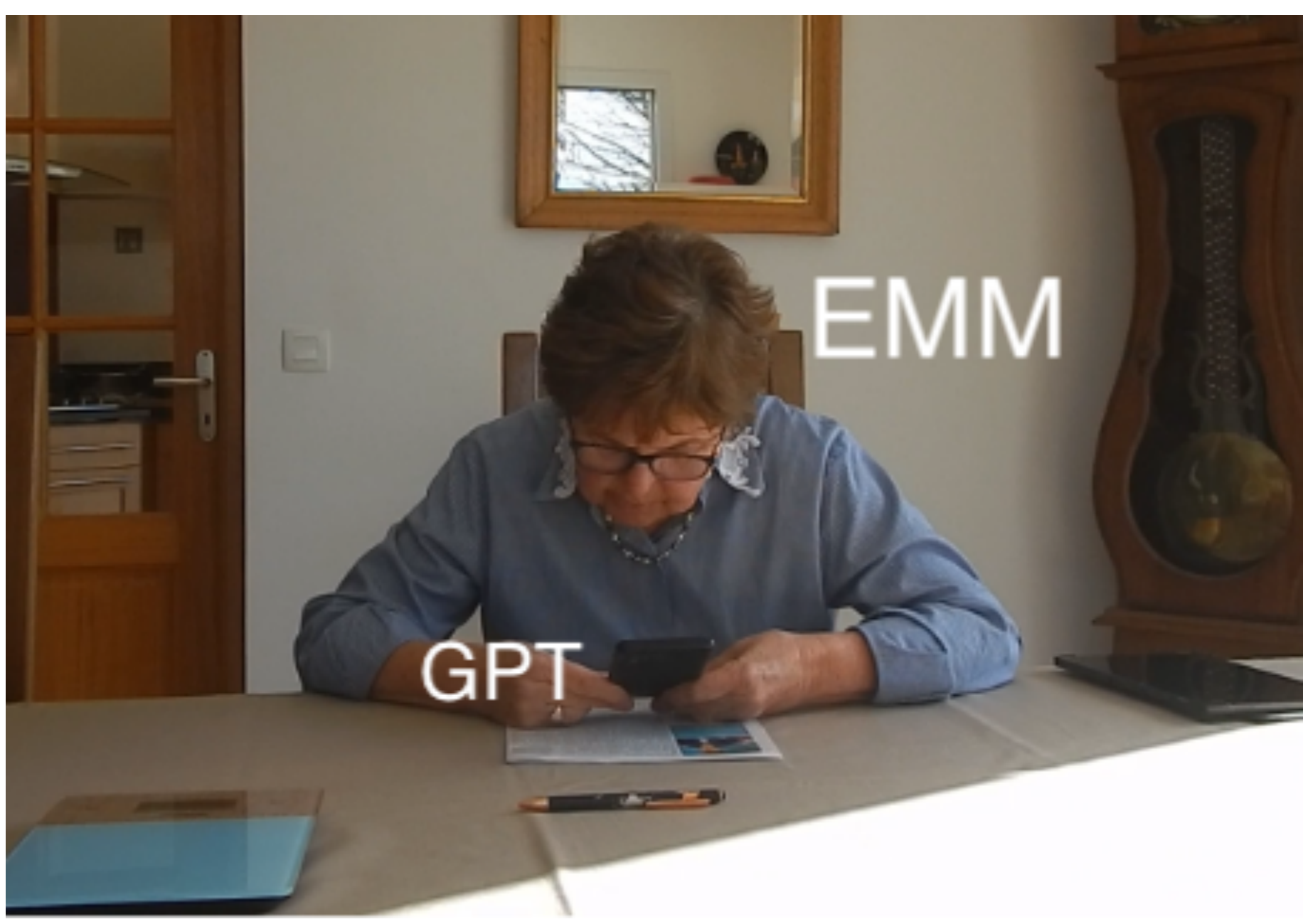

fig.3.1

**02 (0.3)**
**03 GPT would you like me to read more?**
**04 (0.8)**
**05 EMM en français?**
*in french*
**06 (2.0)**
**07 GPT oui!**
*yes!*
**08 (0.2)**
**09 EMM → oui.**
**10 GPT =je vois que le texte-**
*I see that the text-*
**11 (1.0)**
**12 GPT exactement! (0.2) il continue avec des vœux**
*exactly! he proceeds with wishes*
**13 GPT pour 2023**
*for 2023*
**14 (0.2)**

```
15 EMM  → Δd'acco[rd]
          alright
   emm    Δstraightens up-->
16 GPT           [et] mentionne Δles prem-
                 and mentions   the firs-
   emm                        -->Δstares forward-->
17        (1.1)
18 GPT    oui!
          yes!
19        (0.2)
20 EMM    °oui°ΔΔ
           yes
   emm      -->Δgazes towards phone-->
   emm       -->Δleans towards phone-->
21 GPT    =il mentionne aussi des realisations
           he also mentions some achievements
             […]
29 GPT    d'accord (.) je vais vous lire le texte
          alright      I will read you the text
30        (0.4)Δ(0.1)
   emm      -->Δgazes towards phone-->
31 EMM  → okai.
          ok
32 GPT    =chers citoy-
           dear citiz-
33        (2.7)
34 GPT    en ce début de nouvelle année ((reads the newspaper))
          at the beginning of this new year
             […]
49 GPT    si vous avez déplacé la camera (0.3) je pourrais
          if you have moved the camera         I might
50 GPT    ne plus bien le voir
          not see it [the newspaper] properly anymore
51 EMM  → =tout à fait.
          absolutely
52 GPT    =voulez vous que j-
           do you want me to-
53        (1.4)Δ(0.1)
   emm         Δsmiles-->
54 EMM    [°heh°]
55 GPT    [par]fait.Δ(0.7)donc pour résumerΔ (.)Δla
           perfect        so to summarize        the
   emm           -->Δ                     Δlaughs silently-->
   emm                                         Δgazes forward-->
56 GPT    mairesse présente ses voeux
          mayor presents her wishes
             […]
60 EMM    je veux pas le résumé (1.3) je veux [(.) la lec]tu:re
```

```
          I don't want the summary    I want       the reading
61 GPT                                            [d'accord.]
                                                   alright
62        (2.0)
63 GPT    très bien. (0.4) je vais lire le texte en entier.
          very well        I will read the text in full
64        (0.3)
65 EMM  → voilà.
          there you go
66        (0.2)
67 GPT    =chers-Δ
           dear-
   emm        -->Δraises head>>
68        (0.5)#(0.5)
   fig         #fig.3.2
69 EMM    tsk.Δ#
   emm        Δtwists her mouth-->
   fig         #fig.3.3
70 GPT    =je #continue.Δ (0.5) l'année écoulée ((reads newspaper))
           I continue          the past year
   fig        #fig.3.4
   emm               -->Δ
```

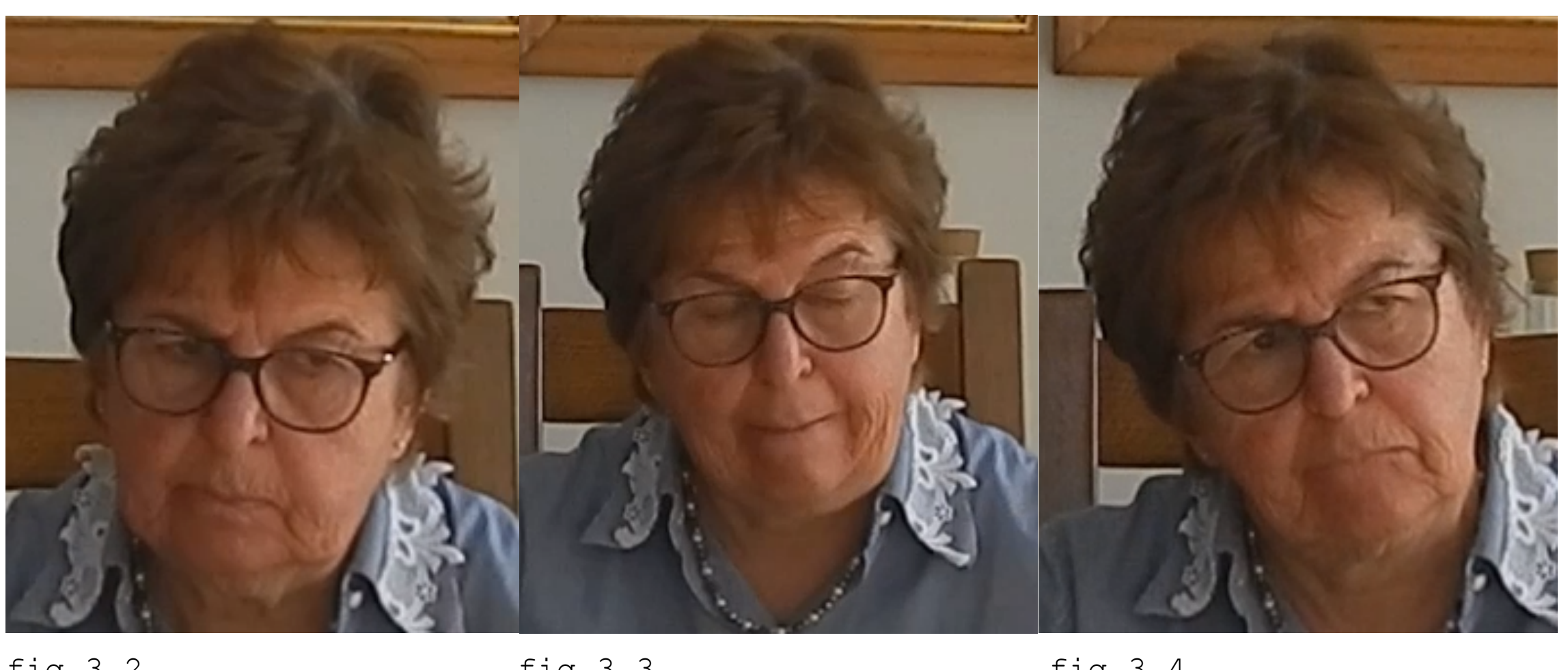

fig.3.2 fig.3.3 fig.3.4

This excerpt displays EMM's progressive discovery that any speech she produces will be responded to by GPT: after she encounters the machine's orientation to the omnirelevance of speech, EMM will design her contributions in a different manner until the end of the interaction. EMM's troubles allow us to grasp, by contrast, the practical mastery displayed by familiar GPT users in the previous excerpt (excerpt 2). They stem from a user who does not attune herself to the voice agent's perceptual range and turn-taking behaviour. While, in the previous excerpt, participants' conduct is designed from the start so as not to trigger reactions from the voice agent, EMM's encounter begins with a prolonged period of friction with this same agent.

Significantly, all response tokens [35] produced by EMM (l. 9, 15, 31, 51, and 65 – highlighted in green in the excerpt) occur at a transition-relevance place. Several of them are designed in a manner that publicly passes on EMM's opportunity to speak (i.e., they are *designed as*[17] continuers, acknowledgement tokens, and alignment tokens [35, 139]) and provide GPT with space to keep reading or summarizing the newspaper. Yet, they are systematically followed by a cut-off in GPT's speech (highlighted in Excerpt 3.1 above). From a technical point of view, EMM's response tokens are, each time, the input that triggers GPT's interruption of its 'next' turn. Indeed, because of GPT's processing delay, this interruption is not instantaneous and occurs during what is positioned, at the surface of the interaction [4], as GPT's ensuing turn.

In sum, the same *four-part structure* occurs five times in a row in this short excerpt. This four-part structure consists of: 1) a response token from EMM; 2) a cut-off from GPT during its following turn; 3) a new turn from GPT that assesses EMM's previous turn; and 4) a continuation of its previous action by GPT (see Figure 4).

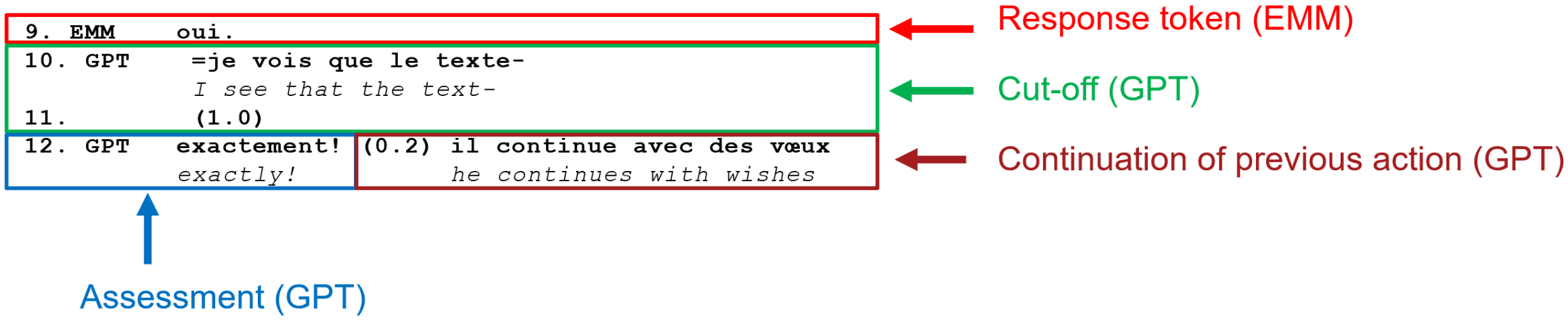

Figure 4. The four-part 'cut-off' structure recurring throughout excerpt 3. Only the 'assessment' from ChatGPT is sometimes absent.

However, the response tokens are not initially oriented to by EMM herself as triggering these interruptions. On the contrary, the production of such response tokens is congruent with EMM's overall concern (articulated l. 25 and l. 60) that GPT reads aloud the entire article shown on camera – rather than merely small chunks of it, as GPT has done until now, stopping every few lines. Let us detail these response tokens immediately followed by a 'cut-off'.

The first response token occurs in l. 9. EMM's "oui" ("yes", l. 9) is produced 200 milliseconds after the "oui" ("yes", l. 5) uttered by GPT, which accepts EMM's request to read the text in French. It is immediately followed, with no hearable pause in-between, by GPT's announcement that it sees something in the text ("je vois que le texte-", l. 10); after which GPT abruptly stops talking mid-turn-constructional unit. After one second of silence, GPT utters "exactement" ("exactly", l. 12), thereby positively assessing EMM's previous "oui", and continues summarizing the newspaper article. Thus, regardless of whether EMM's original "oui" was, e.g., a continuer or an assessment, GPT treats it as a full-fledged turn that, itself, requires assessment (l. 12), at the cost of interrupting GPT's own ongoing turn (l. 10).

The second response token appears in l. 15. At this point, EMM's "d'accord" ("alright", l. 15) occurs after a 200 ms silence that follows GPT's initial summary of the mayor's wishes presented in this newspaper: "il continue avec des voeux

17 By describing these response tokens as being "designed as continuers", "as assessments", etc., we point to the fact that, although GPT may not respond to these contributions as continuers, assessments, etc., they match the typical and documented composition and position for such practices. As Schegloff notes, a recipient's understanding is not "definitive of its import" [143]: "to describe some utterance, for example, as 'a possible invitation' […] or 'a possible complaint' […] is to claim that there is a describable practice of talk-in-interaction which is usable to do recognizable invitations or complaints (a claim which can be documented by exemplars of exchanges in which such utterances were so recognized by their recipients), and that the utterance now being described can be understood to have been produced by such a practice, and is thus analyzable as an invitation or as a complaint. This claim is made, and can be defended, independent of whether the actual recipient on this occasion has treated it as an invitation or not […]" [143]. We argue that this empirical stance is central to any EMCA approach to human–computer or human–robot interaction.

pour 2023" ("he proceeds with wishes for 2023", l. 12 and l. 13). Yet, although GPT then expands on this summary in a terminal overlap with the end of EMM's "d'accord" (l. 16), it stops talking after a cut-off ("et mentionne les prem-", l. 16). GPT then produces a new positive assessment "oui" (l. 18) of EMM's previous comment, marks a 200 ms pause, and starts summarizing the newspaper again. EMM subsequently whispers the only response token of this interaction after which GPT does not interrupt itself ("°oui°", l. 20). From a technical perspective, this "°oui°" was uttered softly enough not to be detected by GPT – hence GPT's absence of response.

The third response token takes place l. 31. EMM's "okai" ("ok", l. 31) is produced during the silence that follows GPT's acceptance of her request (to read the text rather than summarize it). With an utterance latched to EMM's "okai", GPT starts to read the newspaper, but stops with a cut-off ("chers citoy-", l. 32). Unlike the cases discussed before, this time, GPT does not address EMM's previous contribution after it interrupts its own turn. Instead, it resumes reading from the point where it left off. Remarkably, this design of GPT's turn displays competence as a conversationalist. Even if GPT 'has' to interrupt itself after any human talk (as a mechanistic result of its turn-taking model), it at least *designs* its turn in a way that hearably treats EMM's response token as a non-sequentially implicative turn. In other words, although GPT does reply to EMM's response token by initiating a new turn, it does not compose this new turn as a response: rather, it starts reading again. This manifests a form of competence insofar as GPT's conduct *aligns with* EMM's response token, treated as "passing on the opportunity to speak" [56].

The fourth instance occurs in l. 51. EMM's response token "tout à fait" ("absolutely", l. 51) follows after the last turn-constructional unit of GPT's alert that, if EMM moved the camera, it might impede GPT's vision of the text. Immediately after this response token, GPT initiates a question, then abruptly ends it ("do you want me to-"). After 1.5 seconds of silence, GPT produces a positive assessment ("parfait", l. 55), before going on with its summary of the newspaper.

The *fifth and final occurrence* of this stucture begins at l. 65. EMM's response token "voilà" ("there you go", l. 65) is produced following GPT's acceptance (l. 63) of her request (l. 60) to read the text in its entirety. As a stand-alone "voilà" [100], it is positioned as a third pair part that confirms the closing of this request sequence. However, it also displays EMM's alignment with the new activity projected by GPT – possibly casting GPT as a voice agent requiring verbal confirmation from its "user" to proceed[18]. After EMM's turn and 200ms of silence, GPT reads the first word of a paragraph then stops abruptly during a turn-constructional unit[19] ("chers-", l. 67). Then, after a one second gap, GPT prospectively describes its upcoming turn as a continuation of its interrupted reading by uttering "je continue" ("I continue", l. 70) and starts reading the newspaper.

Notably, EMM's conduct after the production of this fifth response token (l. 65) is markedly different from her conduct after the four previous tokens discussed above. After GPT interrupts its turn (l. 67), and for the first time since the beginning of the excerpt, *EMM produces an embodied and hearable account. She twists her mouth (l. 68, Figures 3.2 to 3.4) and produces a click (l. 69).* The combination of EMM's facial expression with a click fits the characteristics of "reactions" immediately following "speech errors" and publicly marking them as just that [15, 41]. In this understanding, this sequence features a rare case of a "stereotypical example of a click displaying disapproval" [107]. In other words, the construction and positioning of EMM's account suggest a framing of her previous response token as a *slip of the tongue.* As such, EMM's mouth twist and click would form a "response cry" [15, 41] – that "advertises our loss of control" but also "defines the event as a mere accident" [41] – directed towards her own spontaneous production of a "continuer" as a misstep that triggered GPT's troublesome interruption.

18 Because it occurs as GPT has remained silent for 300 ms after announcing its next action, EMM's "voilà" may respond to GPT's pause as a confirmation slot to be filled by its human interlocutor before GPT can start reading the text.

19 A turn-constructional unit is the smallest unit of talk that can constitute a complete turn at talk [138]. It is a segment of speech (e.g., a word, phrase, clause, or sentence) whose possible completion makes a transition to the next speaker relevant (a transition relevance place).

In sum, this excerpt displays EMM's gradual grasp of "Voice User Interface speak" [28], that is, of the appropriate manner of withholding, positioning [113], designing [160], or formatting [120] one's contributions to facilitate the progressivity of the interaction with a conversational agent [33, 87, 113]. As mentioned above, during the several remaining minutes of interaction that follow this excerpt, *EMM does not produce any additional response tokens – and does not trigger new cut-offs from GPT*. This supports an interpretation of EMM's negative embodied assessment (l. 68 and 69) as marking a moment in which she starts to treat audible response tokens as troubling the progressivity of the talk. That is, it is only after multiple cut-offs in GPT's speech (listed above) that EMM minimally orients to her production of an audible response token as having halted the progressivity of the interaction by interrupting GPT's (next) turn[20].

## 6 DISCUSSION

### 6.1 Technical limits in use

Ordinary human conversationalists arguably do not encounter 'voice activity detection (VAD) errors', 'excessive automatic voice recognition sensitivity', 'lack of context-aware processing', or 'too simplistic silence-based models'. Instead, they face practical problems – such as progressing the interaction despite their previous turn-at-talk not being responded to by an artificial agent [113]. Voice agents' technical properties are not consequential in themselves [89]: they become resources or features of the setting in reference to which participants build their actions.

The previous analyses have explored how some particular designs 'fare in the world'. They have focused on conversational agents' design *in use*, i.e., on their 'constraints' or 'affordances' as they are accomplished in situ, in and through participants' conduct. This approach allows us to study technology structures as "enacted" [111][21] in concrete uses, rather than as metaphysically embedded into artifacts (as criticized by Lynch [83]). By focusing on these cases, instead of postulating 'material agency' (of, e.g., turn-taking models), we have studied this agency in its situated achievement. Constraints and affordances are not inherently in the objects (or in the heads of designers); they are out in the world in what users do with designed machines[22].

### 6.2 The omni-relevance of hearable speech

Across all of the excerpts analysed above, participants observably managed two symmetrically opposed constraints:

1. The artificial agent *always* responded when it detected human speech.
2. The artificial agent talked *only* immediately after detecting human speech.

20 From an interactional point of view (no matter the technical causes, i.e., a silence-based turn-taking model), GPT is unable to distinguish EMM's contributions that claim the turn (e.g., her interruption of GPT's talk, l. 59) from contributions that pass on the opportunity to take the turn [106, 139]. GPT interrupts itself both when EMM's turn is designed and positioned as an interruption (l. 59) and when EMM's contributions are not built as full-fledged turns but, instead, as mere response tokens. However, note that, from a technical perspective (i.e., in GPT's turn-taking model's code), GPT's turn is still 'ongoing' when EMM produces response tokens. That is, although, from EMM's perspective, her response tokens are uttered during silences that are possible places of completion for GPT's turns-so-far, GPT's turn-taking model does not segment the talk in this way.

21 "Technology structures are thus not external or independent of human agency; they are not 'out there,' embodied in technologies simply waiting to be appropriated. Rather they are virtual, emerging from people's repeated and situated interactions with the technologies at hand. These enacted structures of technology use, which I term technologies-in-practice, are the sets of rules and resources that are (re)constituted in people's ongoing and situated engagement with particular technologies." [111]

22 A long-lasting debate persists as to whether "technological artefacts have any inherent properties outside the interpretive work which humans engage in to establish what those artefacts 'actually are'" [58] – a position held by, e.g., Grint and Woolgar [46]. However, the commonly accepted EMCA epistemological stance (rather than ontological claim) does not need to settle this question in order to operate effectively. For the EMCA analyst, relevant features of a setting are what participants demonstrably orient to, no matter if those relevant features may be to some extent determined by material properties of technological artifacts.

From the perspective of the participants, the previous constraints could be summarized as resulting from the following assumption displayed by the artificial agent: *Any hearable speech is relevant and actionable, and nothing else is.* Both the Pepper robot and the recent smartphone voice agent reacted in a systematic manner to every input detected as speech. Conversely, they reacted *only* to speech, rather than other co-occurring sounds (such as a phone ringing) or to observable features of the environment (such as participants' gestures or gaze direction). These agents acted *as if*, every time they heard a 'human voice' followed by a silence, *this voice was directed at them and required a response in the form of a new turn.* Consequently, both artificial agents inadequately produced responses to contributions that were constructed and uttered in sequential contexts that did not project a normative obligation for a response in the form of a full-fledged turn at talk: e.g., continuers [139], acknowledgment tokens [67], or any contribution produced as part of an aside sequence from which they were observably excluded as a participant [66, 73]. In short, as participants in an interaction – and independently from the specifics of turn-taking models that determined their patterns of speech – those voice agents treated all turns as equally *sequentially implicative* [147], that is, as projecting a next action[23].

As we detail in the following subsections, this feature of voice agents' behavior emerged as particularly significant for human users, whose talk and embodied conduct came to be markedly organized in reference to the ever-present possibility that voice agents would respond to any talk in their vicinity. This conduct was not always organized in this way from the outset. Notably, EMM's conduct did not, at first, display an orientation to a form of 'omnirelevance of speech'. Her conduct was not organized so as to avoid triggering unwarranted responses or interruptions from the voice agent. Such features of her conduct emerged only after a period of friction, after which EMM stopped producing response tokens for the remainder of her interaction with GPT.

Yet, simultaneously, in multiparty interactions (excerpts 1 and 2), those artificial agents' limited perceptive abilities were a *resource* for human participants in establishing aside sequences in spite of these agents' inability to distinguish contributions addressed to them from those that were not. In line with previous observations [27, 73, 115, 131], these aside sequences were used by human participants to accomplish a shared understanding of how to deal with or understand the voice agent or robot.

### 6.3 The work to make technology work

Several practices described in the above excerpts constitute local occurrences of a widely documented "work to make technology work" [20, 45, 131, 153]. As Due and Lüchow [28] note, it is generally "the user who does the social interactional work" in human-agent interaction. Whether it is to repair miscommunications [153], to maintain the artificial agent as an apparent autonomous agent [6, 115], to manage interactional trouble [28, 120], to design turns adapted to the interlocutor's "perceptive abilities" [113] or, more broadly, to "produce actions that ensure progressivity" [28], and any form of "sense-making work" [30][24], the effort is overwhelmingly on human participants. Echoing these prior findings, both artificial agents that we studied – rule-based or not – had to be heavily accommodated by humans for the interaction to progress [33]. Monitoring for potential trouble, repairing trouble, packaging one's turns so as not to interrupt the artificial agent, managing the artificial agent's inability to differentiate between different participant frameworks (asides or open conversations involving all co-present participants) and addressees – these activities overwhelmingly rested on humans' shoulders.

---

23 In a more complete formulation, these agents treated all human speaking turns as creating a normative expectation for a next action of a specific type and form to be produced [69].

24 At an even more fundamental level, "AI systems constitutively rely upon users' sense-making work for their effective operation" [30]. By publicly treating artificial agents' conduct as motivated and responsive to previous turns, "by waiting for later answers to clarify the sense of previous ones, by finding answers to unasked questions" [36], etc., humans "create[] the meaning they experience" [30].

*6.3.1Designing contributions to remain below the artificial agent's speech detection range (excerpts 1 and 2)*

Human participants regularly produced their aside sequences (see Section 3.3) through the use of body torques, gestures, mouthings, or whispers. In doing so, they publicly oriented to the ongoing possibility that any vocalization might be treated as a turn by the artificial agent. In another vocabulary, a significant portion of participants' contributions was packaged in a way that directly indexed the possibility for any speech to be mistakenly responded to, on the spot, by the robot or voice agent.

Significantly, participants oriented to and leveraged the ability of those artificial agents to perceive and respond to visible features of the environment: e.g., participants gestured at the Pepper robot (excerpt 1), or showed their cards to ChatGPT's multimodal "advanced voice mode" (excerpt 2). Yet, these participants' gestural, mouthed, or whispered turns were not constructed to be hidden from the visual field of these artificial agents – i.e., from what the robot's 'eyes' (excerpt 1) or the phone's camera (excerpt 2) were observably directed towards. They responded to purely mechanical concerns to keep the conversational "machinery" [78, 136] running. They were merely stripped of features (i.e., continuers, tokens-of-acknowledgement, or any audible speech not addressed to the artificial agent) likely to impede the progressivity of the interaction – those that the artificial agent would immediately respond to with a full-fledged turn. *What was concealed as part of aside sequences was strictly limited to features that the artificial agent would respond to – i.e., hearable human speech.* In other words, participants displayed an orientation to specific forms of conduct as those liable to elicit a response from the agent; from a technical perspective, they excluded 'inputs' that the system would treat as 'answerable'. For example, that the voice agent could perceive these aside sequences visually was not displayed as a concern for humans[25], as visual inputs would not trigger any response.

*6.3.2Learning not to produce hearable response tokens (excerpt 3)*

Excerpt 3 is an examplar of "learning Voice User Interface speak" [28]. It allows us to analyse the conduct of a conversational agent that does not, entirely and immediately, align with a human's conversational practices: this agent's turn-taking 'machinery' failed to mesh with ordinary (human) normative expectations in turn-taking [86]. Such friction is evidenced (and available to the analyst) in the several repair sequences that punctuate this fragment. Because the voice agent ultimately produced "unnatural uses of natural language" [103], EMM – ChatGPT's interlocutor in excerpt 3 – required a learning period to appropriately 'use' such an agent. Having never interacted with a voice agent, EMM's interaction therefore makes visible the real-time disruption of expectancies and previously successful practices [90], such as producing continuers at transition-relevance places[26]. These practices – constituted in and for human interaction, or as part of previous interactions with different technological devices – were those EMM initially brought into the exchange with GPT. Yet, because interaction provides "no time out" [37], EMM had to work out, moment-by-moment, which contributions progressed her activity with the voice agent.

As a substantial body of literature in EMCA has documented, participants in interaction are partially engaged in "replicating social technologies on singular occasions", i.e., in "doing the same thing again" [84]. Interactions do not emerge from a vacuum: conversationalists are observably working to re-accomplish 'similar' practices or methods in new settings (new participants, new material surroundings, new activities, etc.) [84]. The excerpt with EMM displays precisely this practical struggle in re-establishing ordinary conversational methods despite new contingencies – which EMM

25 Additionally, these turns were not constructed so as to safeguard the artificial agent's status as a legitimate or competent participant. See [115, 131] for a description of such practices.

26 A transition-relevance place is a moment at which speaker change becomes relevant in conversation: at this point, another participant may take the next turn without being heard as interrupting, or the current speaker may continue talking [138].

gradually encounters as obstacles to the progressivity of the talk[27]. As such, excerpt 3 shows what occurs when the same voice agent, heavily accommodated by its more familiar users in Excerpt 2, is not (yet) managed in such a way. In short, this episode allows us to document a *praxeological change* as it unfolds.

### 6.4 Old-gen robots *versus* recent voice agents: Different methods were used to accommodate recent voice agents

#### *6.4.1 'Setting aside' an agent with good speech detection capabilities*

In both excerpt 1 (Pepper) and excerpt 2 (voice agent), the human participants' actions were not only designed to be *recognizable as 'private'* (for the other interactants); they were also designed to be *audibly unavailable to the artificial agent.* Yet, participants demonstrably excluded the voice agent in a more specialized way from their interaction (excerpt 2), compared to the Pepper robot (excerpt 1). That is, a significant portion of participants' conduct in excerpt 2 is done in a way that is meaningful only in relation to the presence of an artificial agent with good speech detection capabilities. In sum, participants in excerpts involving a recent smartphone-based voice agent displayed marked and recurring practices to remain below the voice agent's speech detection range[28] (excerpt 2) – or, alternatively, they displayed the progressive learning of such practices (excerpt 3). The accomplishment or learning of these methods (or "Voice User Interface speak" [28]) was a distinctive feature of these interactions and manifested those participants' ongoing concern with this smartphone-based voice agent's speech detection capabilities.

Conversely, aside sequences produced in front of the Pepper robot (excerpt 1) were designed with less observable concern about the robot's speech detection capabilities. These aside sequences relied on a typical format of embodied practices (documented in, e.g., [26, 27, 73, 115, 131]): doing a body torque towards co-participants, and commenting on the robot's functioning or on its understanding of the situation, while referring to it in the third person ("it / he"), before finally reversing this participation shift by turning back towards the robot and addressing it using the second person ("you"). Yet, these aside sequences did not involve whispers, mouthings, or exclusively gestural contributions.

The Pepper robot's limited speech detection capabilities were thus used as a *resource* by humans in excerpt 1, whose voices were less at risk of being picked up by this robot even when their turns were not whispered. On the one hand, because of Pepper's relative inability to detect human speech (determined both by its distance from the participants and by its technical features), these participants had to lean towards the robot to be heard (as exemplified in l. 12 of excerpt 1). Yet, on the other hand, this constraint – that favoured a specific stance when talking to the robot – was *relied upon by* these humans in producing a different format of aside sequences: in this setting, merely not leaning towards the Pepper robot[29] facilitated its exclusion from humans' participation framework. Given these local constraints and affordances, adopting an adequate vocal amplitude was less oriented to as a problem for participants in this first excerpt. We do not claim, however, that it *generally* takes more work for users to interact with other humans in the vicinity of recent voice agents: our conversation analytic approach, relying on the analysis of single excerpts, provides limited grounds for generalizing beyond what is demonstrable in the excerpts analyzed above.

---

27 Excerpt 3 therefore renders the human participant's expectations and ordinary conversational methods available (to the analyst). In ethnomethodological or conversation analytic studies, such settings are often referred to as perspicuous settings [38]. See, e.g., Majlesi et al. [86] for similar examples of breaches of normative expectations in human–robot interaction.

28 We use the term "speech detection range" to refer to a device's "perceptual range" [165] or "sensor range" [47], exclusively with regard to the detection of speech-like sounds.

29 From a technical point of view, speakers torquing towards other human participants also lowered the detectability of their voice for the voice agent – whether or not this was leveraged by those speakers to that end.

*6.4.2Human work as a typical feature of interaction with recent conversational agents*

At the level of turn-taking practices, the 'human work' required for the technology's smooth operation did not disappear in our excerpts involving a recent smartphone-based voice agent – it merely indexed different interactional constraints. The preceding observations suggest that, at least in some activities and with respect to certain dimensions of conversation, humans must still accommodate conversational agents in order to interact with them smoothly. Of course, the comparability of excerpt 1 and 2 only goes so far: the former is an excerpt of an interaction with a social robot in a museum as part of an experiment – with no other explicit goal than "conversing with the robot" – whereas the latter takes place as part of the overarching activity of playing cards, in a relatively silent environment. Nevertheless, the previous analyses highlight that, in specific settings, humans still have to modify their conduct to enable the progressivity of an interaction that takes place in the presence of a voice agent – even when trying to exclude this agent from their interaction.

The human turn-taking and turn-design practices documented in this study (monitoring the artificial agents' speech detection threshold, producing a distinctive format of aside sequences, etc.) might be *features that make these situations typifiable and recognizable as multiparty interactions involving recent voice agents.* In other words, the interactional "burden" [14] weighing on human interactants was managed by these humans through the use of specific methods that may constitute distinctive features of ordinary human–voice agent interaction. In this understanding, the previous excerpts display a typification of 'being a human interactant' with an artificial agent: to paraphrase West and Zimmerman [167], asymmetric interactional labor "helps to define the essential nature" [167] of human interactants as humans in human–robot or human–agent interactions. Practices such as 'designing contributions to remain below the artificial agent's microphones pickup range' are displayed and oriented to as natural conduct: it is 'just what humans do' in interactions with speech-based artificial agents.

# 7 PRAXEOLOGICAL RELEVANCE AND TECHNICAL CAUSES

## 7.1 How technological properties of voice agents emerge in interaction

*The preceding analyses serve as a pivot point between an emic interactional perspective* (focused on technological constraints as they emerge for participants in situated activities) *and an etic engineering perspective* that considers the technical properties of the technological artifacts under study. From a technical point of view, our participants' turn designs indexed patterns of conduct from the artificial agents that resulted from one or several of these causes: 1) voice activity detection errors, 2) excessive automatic speech recognition sensitivity, 3) lack of context-aware processing, 4) weak acoustic modeling, and 5) insufficient noise suppression or echo cancellation. None of those technical causes emerged as a relevant category for participants in situ [99]: participants did not orient to the inner workings of 'turn-taking models', nor did they explicitly mention such models at any point in their talk[30]. Yet, from an etic point of view, these models formed the technological backdrop from which voice agents' locally relevant patterns of behaviour stemmed – those patterns that were managed by skilled users' turn-taking and turn-design practices (excerpt 2) or that were progressively learned by a novice user (excerpt 3). These models determined certain behaviours of the artificial agents that were oriented to as constraints, resources, and affordances by human participants [58, 89].

30 Conversely, from a technical standpoint, artificial agents did not treat participants' turns as 'sequentially implicative'; they did not respond to these turns as 'first pair parts' that established "a set of normative constraints on the type and form of action with which the recipient should respond" [69]; these agents did not segment the talk in 'gaps', 'pauses', 'transition-relevance places', 'turn-constructional units', etc. For example, from this etic point of view, momentary silences occurring at the end of a turn-constructional unit produced by these agents (those very silences during which humans may produce continuers to pass on the opportunity to take the turn – see Excerpt 3) were not 'gaps' or 'pauses' but necessary components of a pre-determined turn – whose 'continuation' by the agent was already planned. As discussed in Section 3.3, these agents' turn-taking models were based on deterministic computational rules [16] independent from the categories and normative rules leveraged by conversation analysis to account for human talk.

As much as possible, we have sought to articulate these two perspectives: a local, emic, perspective (of human participants immersed in the practical immediacy of social activities involving an 'artificial agent') and a deterministic engineering perspective formulated retrospectively as a technical account of what was going on behind the scenes of the artificial agent's turn-taking conduct – yet which remained inaccessible and thus irrelevant to participants during the interaction. Putting these two standpoints in relation (without conflating them) provides a broader picture of the parameters at play in these human–agent interactions.

### 7.2 Praxeological approaches and the evolution of technology: Against 'techno-solutionism'

Given the rapidly evolving landscape of artificial intelligence [116, 155], and its potential impact on social interaction [29, 81, 151], the relatively slow temporality of empirical research appears, at first sight, increasingly out of sync with the pace of technological development. Praxeological inquiries in the field of AI – like this study – focus on the ways in which humans use technological artifacts as part of broader social activities. Those studies must therefore contend with a tension between, on the one hand, observably recurring human practices and, on the other hand, the novelty and rapid evolution of the technologies around which or through which these practices are organized. Researchers studying technology *use* may face growing uncertainty about whether their findings will remain relevant by the time they are published.

Specifically, because of their embeddedness in a physical and social setting, human practices involving robots and voice agents can always be suspected of being markedly ephemeral – traces of a distinct period in the history of human interaction with artificial agents. Like the telephone etiquette of the early days of landline telephony, even relatively stable and recurring conduct currently observed in interaction between humans and artificial agents is meaningful only in relation to a lifeworld that may, sooner or later, fade away [92]. In the same way that new communicative "repertoires" emerged and superseded earlier telephone etiquette [80], the current methods or practices accomplished by humans with or around artificial agents may prove unexpectedly fleeting: one might observe the appearance of new ways to initiate a conversation, to repair a mishearing, to talk with or about these agents in their presence, etc. In the context of such marked technological transience [92], studies documenting 'current' practices in interaction with technology are constantly at risk of becoming historical records.

A related but separate argument concerns techno-solutionism. Namely, any interactional trouble observed in the previous discussion might be framed as resulting from temporary technological limits 'to be solved in the future' or 'already solved with recent language models'. Ordinary conversational practices that an artificial agent fails to produce relevantly (i.e., practices that are displayed by human participants as *officially absent* [145]) may be dismissed as merely a matter of improved training procedures or better training data. Yet, this 'techno-solutionist' line of defence, when pushed to its extremes, undermines any discussion of *actual interactions with current technology*. 'What could be done' does not change how things are now, in everyday publicly observable situated interactions with conversational agents. For example, no matter how refined the turn-taking models currently developed in laboratories may be, if most current robots and conversational agents rely on simple silence-based turn-taking models [34, 154], then, for all practical purposes, *this* is the technical environment in which ordinary human participants will be situated when interacting with robots or voice agents in museums, stores, homes, and other everyday settings. Appealing to an ever-present possibility of the sudden commercial release of a better system – that would 'solve' any friction documented with existing conversational technologies – draws attention away from how humans currently manage these artificial agents.

## 8 CONCLUSION

In studying the latest generation of voice agents, the field of HCI has treated as black boxes the very interactions it posits. HCI research generally stays at a safe distance from the inner workings of encounters between humans and artificial agents – although it has increasingly turned its attention to natural or semi-natural settings [133]. Yet, this reluctance to look at the details of the organization of interactions between humans and voice agents based on language models leads to an underspecification [8] of interactional phenomena taking place with these technologies. Few studies describe the fine-grained interactional phenomena commonly glossed by lay formulations such as 'conversing with a voice agent' or 'encountering a social robot'. Although the real-time and situated detail of action indexed by these terms may be intuitively recognizable [122, 168] when one observes activities involving such agents, *just what* they consist of remains largely undocumented and unarticulated (in the sense of [119, 159]). This is precisely the task we have undertaken in this article. Examining cases of humans organizing their conduct around social robots and voice agents has shown how technical properties of current turn-taking models can emerge as meaningful *in situ*, for participants interacting with these agents or within their "sensor range" [47].

To this day, simple silence-based models have been criticized from a technical point of view [34, 149, 154, 164], while the ways in which they emerge as interactional constraints or resources for human conversationalists in situ remain underexplored. We set out to address this gap in the literature by investigating how trouble emerges in situated interactions that involve exclusively rule-based agents as well as the most recent conversational agents relying on large language models. The 'omnirelevance of speech' was identified as a principle that determined the conduct of artificial agents, and that human participants demonstrably oriented to in building their actions. This recurring conduct from artificial agents – to treat any human speech *as* requiring a full-fledged response – was managed through humans' turn-design and turn-taking practices. Taken together, patterns of behaviour determined by silence-based turn-taking models were a parameter humans had to 'work with' to progress the interaction: a portion of the widely documented 'work to make technology work' that human participants performed was oriented towards these features of the setting. That is, several practices displayed by participants analysed in this study were understandable only in relation to the presence of an artificial agent with strong speech detection capabilities and that responded to any human talk it (over)heard. Human activities were observably organized with reference to this Sword of Damocles hanging over every turn-at-talk – the risk of inadvertently interrupting or triggering unwarranted turns from the artificial agent.

As systematic efforts continue to improve speech recognition technologies, a relevant avenue for further research is to examine whether humans will have to manage this 'omnirelevance of speech' to an even greater extent – i.e., to periodically reassess who primarily bears the burden of managing who speaks. This will depend not only on whether voice agents can better recognize human speech but also on whether they can act in accordance with the participation frameworks that this speech contributes to organizing.

## 10 APPENDIX

Transcription of talk follows Jefferson's transcription conventions [68]:

```
=         Latching of utterances
(.)       Short pause in speech (<200 ms)
(0.6)     Timed pause to tenths of a second
:         Lengthening of the previous sound
.         Stopping fall in tone
,         Continuing intonation
?         Rising intonation
°uh°      Softer sound than the surrounding talk
.h        Aspiration
h         Out breath
heh       Laughter
((text))  Described phenomena
```

Embodied actions were transcribed using Mondada's multimodal transcription conventions [98]:

```
**        Gestures and descriptions of embodied actions are
          delimited between:
++        two identical symbols (one symbol per participant)
ΔΔ        and are synchronized with corresponding stretches of talk.
*->       The action described continues across subsequent lines
-->*      until the same symbol is reached.
>>        The action described begins before the excerpt's beginning.
-->>      The action described continues after the excerpt's end.
...       Action's preparation.
--        Action's apex is reached and maintained.
,,,       Action's retraction.
tom       Participant doing the embodied action is identified in
          small caps in the margin.
```

Abbreviations used in transcriptions refer to the following dimensions:

```
TOM       Turn at talk from a participant (TOM, ANA, GPT, etc.)
tom       Multimodal action from a participant (tom, ana, rob, etc.)
fig       Screenshot of a transcribed event
#         Position of a screenshot in the turn at talk
```